\documentclass[onecolumn,superscriptaddress,showkeys,preprint]{revtex4-1}
\usepackage{color}
\usepackage{graphicx}
\usepackage{dcolumn}
\usepackage{bm}
\usepackage{SIunits}
\usepackage[unicode=true, bookmarks=true, bookmarksnumbered=false, bookmarksopen=false, breaklinks=false, pdfborder={0 0 1}, backref=false, colorlinks=false]{hyperref}
\hypersetup{pdftitle={Strong Coupling of Intersubband Resonance in a High Electron Mobility Transistor Structure to a THz Metamaterial by Ultrawide Electrical Tuning}, pdfauthor={S. Pal, H. Nong, S. Markmann, N. Kukharchyk, S. R. Valentin, S. Scholz, A. Ludwig, C. Bock, U. Kunze, A. D. Wieck,  N. Jukam}, pdfkeywords={THz metamaterial, intersubband resonance, 2DEG, HEMT, epitaxial gates, FTIR transmission spectroscopy, THz time domain spectroscopy, strong coupling}}

\begin{document}

\title{Strong Coupling of Intersubband Resonance in a High Electron Mobility Transistor Structure to a THz Metamaterial by Ultrawide Electrical Tuning}

\author{Shovon Pal}
\email{shovon.pal@ruhr-uni-bochum.de}
\affiliation{Lehrstuhl f\"{u}r Angewandte Festk\"{o}rperphysik, Ruhr-Universit\"{a}t Bochum, D-44780 Bochum, Germany.}
\affiliation{AG Terahertz Spektroskopie und Technologie, Ruhr-Universit\"{a}t Bochum, D-44780 Bochum, Germany.}

\author{Hanond Nong}
\affiliation{AG Terahertz Spektroskopie und Technologie, Ruhr-Universit\"{a}t Bochum, D-44780 Bochum, Germany.}

\author{Sergej Markmann}
\affiliation{AG Terahertz Spektroskopie und Technologie, Ruhr-Universit\"{a}t Bochum, D-44780 Bochum, Germany.}

\author{Nadezhda Kukharchyk}
\affiliation{Lehrstuhl f\"{u}r Angewandte Festk\"{o}rperphysik, Ruhr-Universit\"{a}t Bochum, D-44780 Bochum, Germany.}

\author{Sascha R. Valentin}
\affiliation{Lehrstuhl f\"{u}r Angewandte Festk\"{o}rperphysik, Ruhr-Universit\"{a}t Bochum, D-44780 Bochum, Germany.}

\author{Sven Scholz}
\affiliation{Lehrstuhl f\"{u}r Angewandte Festk\"{o}rperphysik, Ruhr-Universit\"{a}t Bochum, D-44780 Bochum, Germany.}

\author{Arne Ludwig}
\affiliation{Lehrstuhl f\"{u}r Angewandte Festk\"{o}rperphysik, Ruhr-Universit\"{a}t Bochum, D-44780 Bochum, Germany.}

\author{Claudia Bock}
\affiliation{Lehrstuhl f\"{u}r Werkstoffe und Nanoelektronik, Ruhr-Universit\"{a}t Bochum, D-44780 Bochum, Germany.}

\author{Ulrich Kunze}
\affiliation{Lehrstuhl f\"{u}r Werkstoffe und Nanoelektronik, Ruhr-Universit\"{a}t Bochum, D-44780 Bochum, Germany.}

\author{Andreas D. Wieck}
\affiliation{Lehrstuhl f\"{u}r Angewandte Festk\"{o}rperphysik, Ruhr-Universit\"{a}t Bochum, D-44780 Bochum, Germany.}

\author{Nathan Jukam}
\email{nathan.jukam@ruhr-uni-bochum.de}
\affiliation{AG Terahertz Spektroskopie und Technologie, Ruhr-Universit\"{a}t Bochum, D-44780 Bochum, Germany.}

\date{\today}

\begin{abstract}
\noindent{The interaction between intersubband resonances (ISRs) and metamaterial microcavities can form a strongly coupled system where new resonances form that depend on the coupling strength. Here we present experimental evidence of strong coupling between the cavity resonance of a THz metamaterial and the ISR in a high electron mobility transistor structure with a triangular confinement. The device is electrically switched from an uncoupled to a strongly coupled regime by tuning the ISR with epitaxially grown transparent gates. An asymmetric triangular potential in the heterostructure enables ultrawide electrical tuning of ISR which is an order of magnitude higher as compared to the equivalent square well. For a single triangular well, we achieve a coupling strength of 0.52 THz, with a normalized coupling ratio of 0.26.}
\end{abstract}

\keywords{THz metamaterial, 2DEG, intersubband resonance, HEMT, epitaxial gates, strong coupling}

\maketitle

Ultrastrong light-matter interaction is one of the key aspects of cavity quantum electrodynamics (QED) and quantum photonics and has been a subject of great interest for superconducting qubits \cite{Wallraff2004n,Macha2014nc}, atomic \cite{Kimble1999prl,Kimble2008n,Mckeever2003n} and quantum dot \cite{Hennessy2007n} systems. In addition to electronic and optical frequencies, strongly coupled light-matter systems can also be obtained for quantum well intersubband resonators coupled to mid-infrared \cite{Dini2003prl} and THz \cite{Todorov2009prl,Todorov2010prl} frequencies. When a transition is strongly coupled to a cavity resonance, the bare frequencies of the uncoupled system shift to new frequencies. The frequency shift depends on the strength of the coupling and can be explained in terms of periodic exchange of energy through vacuum Rabi oscillations \cite{Agarwal1984prl,Agarwal1985josab} or the coupling between the oscillators in a strongly dispersive system \cite{Zhu1990prl}. Since frequency shift is proportional to light-matter coupling, switchable and tunable strong coupling is of interest for filters and modulators. Over the years, artificial structures of sub-wavelength sizes with novel electromagnetic properties, commonly known as metamaterials (MMs) \cite{Meyrath2007prb}, are used to confine and enhance the free-space electromagnetic radiation in the near field and hence to study the light-matter interactions in both strong and ultra-strong regimes.

Recently, ultrastrong coupling experiments have been performed, where the cyclotron resonance of the two dimensional electron gas (2DEG) is coupled to the THz split ring resonators (SRRs) \cite{Scalari2012sc,Scalari2013jap}, complementary SRRs \cite{Maissen2014prb} and superconducting complementary metasurfaces \cite{Scalari2014njp} by externally tuning the magnetic field. However, electrical tunability is more preferable for extending the potentials of light-matter coupling into realizable devices. In 2012, Gabbay et al.~\cite{Gabbay2012oe} have modelled the coupling of metamaterials to electrically tunable ISRs in a square quantum well for mid-infrared frequencies, later demonstrated experimentally by Benz et al. \cite{Benz2013oe,Benz2013apl,Benz2013nc}. Due to local bending of the incident electric field in a direction parallel to the growth axis (polarization selection rule) around the metamaterials, they can be employed to couple the free-space THz radiation to excite the intersubband transitions in parabolic quantum wells \cite{Geiser2010apl,Dietze2011oe,Geiser2012apl} and modulation doped multi-quantum wells \cite{Dietze2013apl}. High electron mobility transistor (HEMT) structures, on the other hand, form the building blocks of most modern high-speed electronic circuits. Monolithic integration of MMs with HEMT permits amplitude modulation of THz radiation at MM resonances up to few MHz \cite{Shrekenhamer2011oe} in the linear light-matter regime. Such integrated devices not only show potential application towards electrically tunable THz devices, but also in the ultrastrong light-matter interaction regime.

In this contribution, we exploit the wide tuning possibility of ISR in a single triangular quantum well by an external bias in contrast to the square well, by driving the device from an uncoupled to a strongly coupled regime. We present experimental results on strong coupling of the 2DEG ISR in a modulation-doped HEMT structure to the fixed cavity resonance of a THz MM by electrically tuning the ISR with a high quality epitaxial, complementary-doped and transparent electrostatic gate \cite{Pal2014jpcm}. The tuning mechanism is attributed to the quantum-confined Stark effect \cite{Bastard1983prb}. The use of complementary epitaxial gates is advantageous since the low vacuum band-offset leads to greater electrical tunability.\\

\noindent{{\bf Designing the Metamaterial-HEMT Device}}\\
\begin{figure}[t!]
\includegraphics[width=0.95\columnwidth]{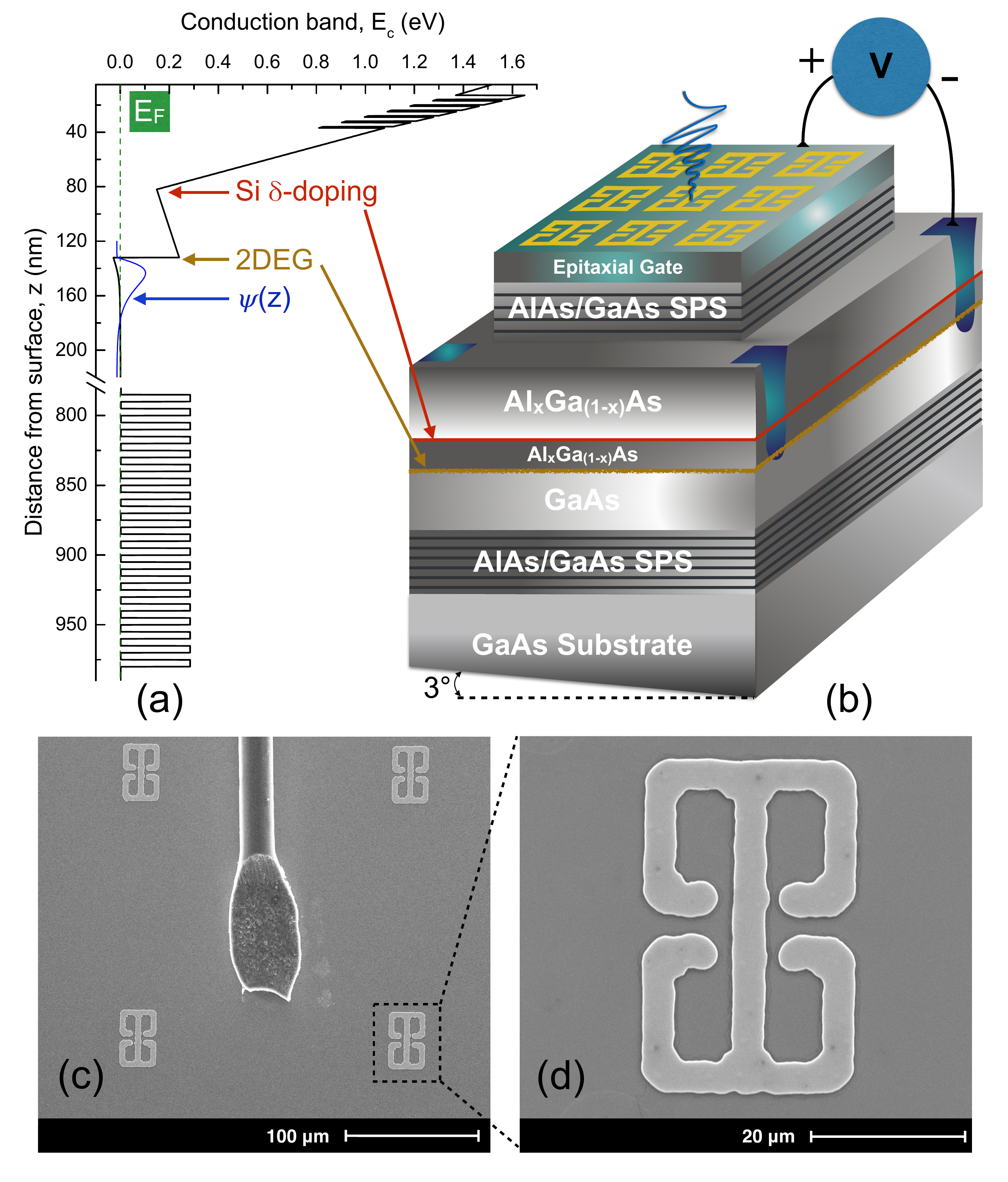}\vspace{-10pt}
\caption{{\bf{Device design.}} \textbf{(a)} Conduction band edge, E$_c$, of our device when a certain bias is applied at the gate such that the ground state is below the Fermi energy. The ground wavefunction is also plotted within the well. \textbf{(b)} Schematic layer sequence of the device and the corresponding electric field direction of the incident infrared radiation which couples to the resonator. The bottom surface of the device is wedged at an angle of 3$^\circ$ in order to avoid the Fabry P\'erot fringes in transmission measurements. SEM micrograph showing \textbf{(c)} metamaterials and wire bonding, \textbf{(d)} metamaterial unit cell.}
\label{fig:device}\vspace{-10pt}
\end{figure}
Intersubband transitions in semiconductor quantum well can be designed to cover the wide infrared region of the electromagnetic spectrum \cite{Craig1996prl,Machhadani2011prb}. The device is based on intersubband transition in 2DEG, formed in a GaAs/Al$_x$Ga$_{1-x}$As heterojunction of a HEMT structure, fig.~\ref{fig:device}(a). The HEMT consists of a heterostructure that is delta-doped in the Al$_x$Ga$_{1-x}$As layer, 50 nm from the heterojunction. The electrons from the delta-doped donors diffuse to the GaAs layer. This results in the presence of a strong built-in space charge fields between the carriers in the GaAs and donors in the Al$_x$Ga$_{1-x}$As. The potential near the heterojunction can be approximated as a triangular potential with a constant slope in GaAs layer from the space-charge field and a vertical barrier from the GaAs/Al$_x$Ga$_{1-x}$As conduction band-offset. The electronic motion perpendicular to the surface is quantized, resulting in the formation of quasi-two-dimensional-subbands. By varying the external bias, the slope of the triangular potential along with the transitions energies can be tuned.  In contrast to a square with abrupt barriers on either side, a triangular well with gradual slope (on the GaAs side) will have a much greater effect on tuning the subband spacings by electrical bias. The external bias is applied by making contacts to the 2DEG and a heavily doped p-type region on the surface which forms an epitaxial gate. The complementary doping of the epitaxial gate leads to a low vacuum band-offset (typically 0.5 eV), thus a lower depletion of carriers and hence a wide voltage tunability of the device (or intersubband levels). The thickness of the epitaxial gate is only 20nm which is much less than the skin depth. This makes the epitaxial gate transparent to THz radiation. Further details on the layer sequence of the sample are described in the methods section.

Arrays of double-sided Au split-ring resonators (SRR) are deposited on top of the device, shown in fig.~\ref{fig:device}(b) and (c). The SRRs have sub-wavelength sizes and are designed to have a cavity resonance comparable to the first intersubband transition in the 2DEG. The resonators are similar to electrical-inductor-capacitor (ELC) circuit, where the inductive regions correspond to the loops of the SRRs and the capacitive regions correspond to the gaps in the loops of the SRRs. As shown in fig.~\ref{fig:device}(d), the double-sided SRRs consists of two single SRRs that are mirror images of each other for reflections along the central metal line \cite{Scalari2012sc}. The current travels in a clockwise direction for one of the single SRRs and in anticlockwise for the other. Thus the induced magnetic fields from the two SSRs will average to zero and the double SRR will only couple to external electric fields through the capacitive regions.

\begin{figure}[t!h]
\includegraphics[width=0.6\columnwidth]{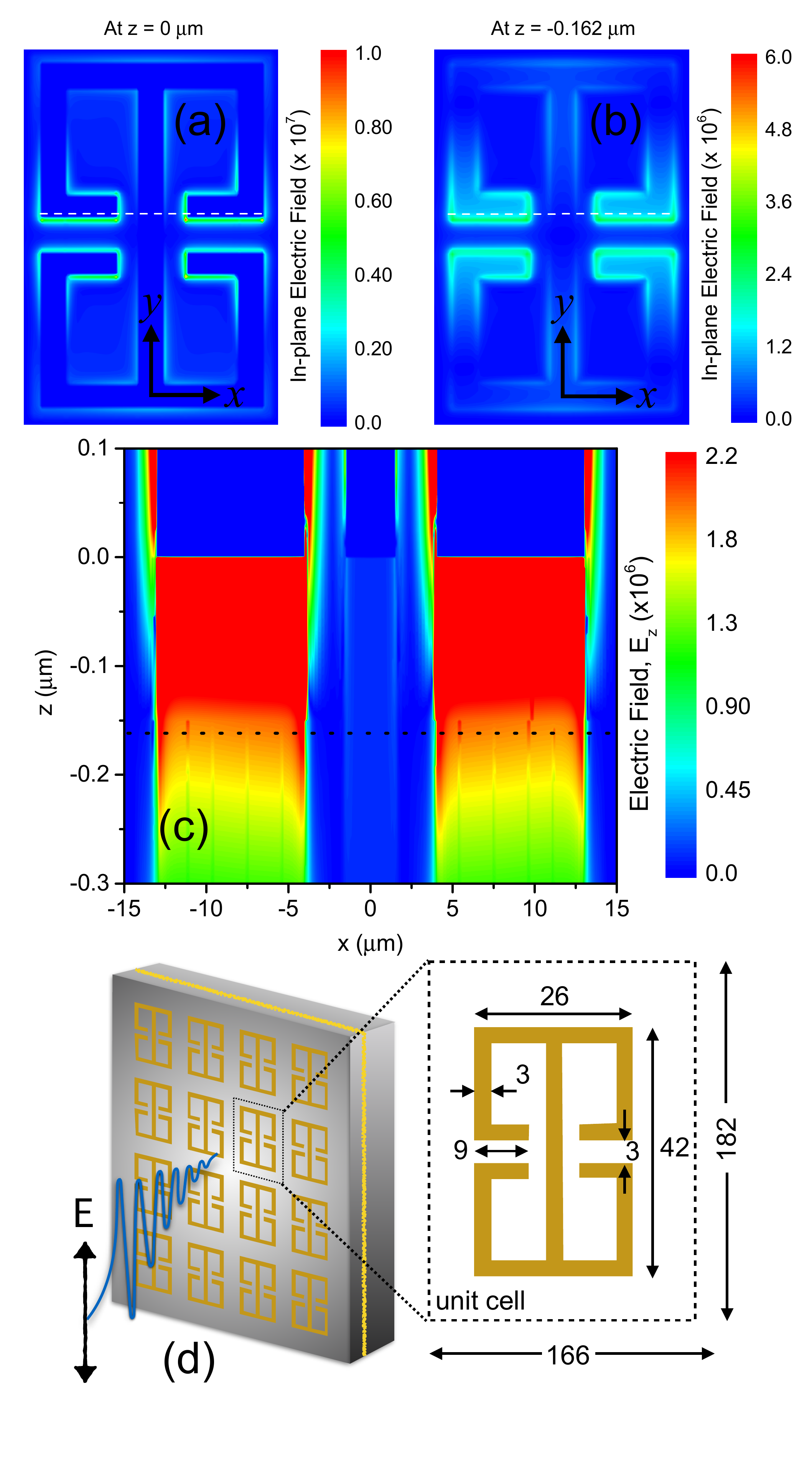}\vspace{-30pt}
\caption{{\bf{FDTD simulations of metamaterials.}} In-plane electric field distribution, ($\sqrt {{{\left| {{E_x}} \right|}^2} + {{\left| {{E_y}} \right|}^2}}$), at \textbf{(a)} $z = 0$ $\micro$m (on top of the surface) and \textbf{(b)} $z = -0.162$ $\micro$m (in the 2DEG layer, 162 nm below the surface). \textbf{(c)} The field distribution, $\left| {{E_z}} \right|$, in the growth direction along the cut shown by the white lines in (a) and (b). The black dotted line represents the position of the 2DEG. \textbf{(d)} The schematic representation of the metamaterial structure and the electric field component of the polarized excitation source for the TDS measurements. All dimensions are in the units of $\micro$m. The metamaterials are arranged in an unit cell of 166 $\micro$m $\times$ 182 $\micro$m.}\vspace{-10pt}
\label{fig:simulation}
\end{figure}

Finite difference time domain (FDTD) simulations using a commercial software package \cite{cst} is used to calculate the electric field distribution and the frequency response of the MM modes. The simulations are carried out for a single unit cell with periodic boundary condition in $x-y$ plane and open boundaries (with spaces) in $z$ plane. The electric field direction of the incident signal is polarized along $y$ direction. A schematic of the arrangement used for electromagnetic simulation along with the dimensions of MM are shown in fig.~\ref{fig:simulation}(d). The in-plane field distribution ${\left (\sqrt {{{\left| {{E_x}} \right|}^2} + {{\left| {{E_y}} \right|}^2}}  \right)}$ is strongly enhanced over sub-wavelength volumes around the capacitive regions of the double-sided SRRs, fig.~\ref{fig:simulation}(a) and (b). Intersubband transitions only couple to electric fields polarized in the growth direction perpendicular to the surface. Although the fields between the capacitor sections are predominantly in-plane, there is a significant out-of plane field from fringing effects on the surface \cite{Shelton2010oe}. This can be seed in fig.~\ref{fig:simulation}(c), where the electric field distribution, ${\left( {\left| {{E_z}} \right|} \right)}$ in the growth direction along the cut, indicated by the white line in fig.~\ref{fig:simulation}(a) and (b) is shown. The equivalent electrostatic capacitance is an important parameter of the microcavities which play a key role in strong light-matter coupling. This is due to the fact that the Rabi frequency is directly proportional to the equivalent capacitance \cite{Benz2015nl}.\\

\begin{figure*}[t!h]
\includegraphics[width=\textwidth]{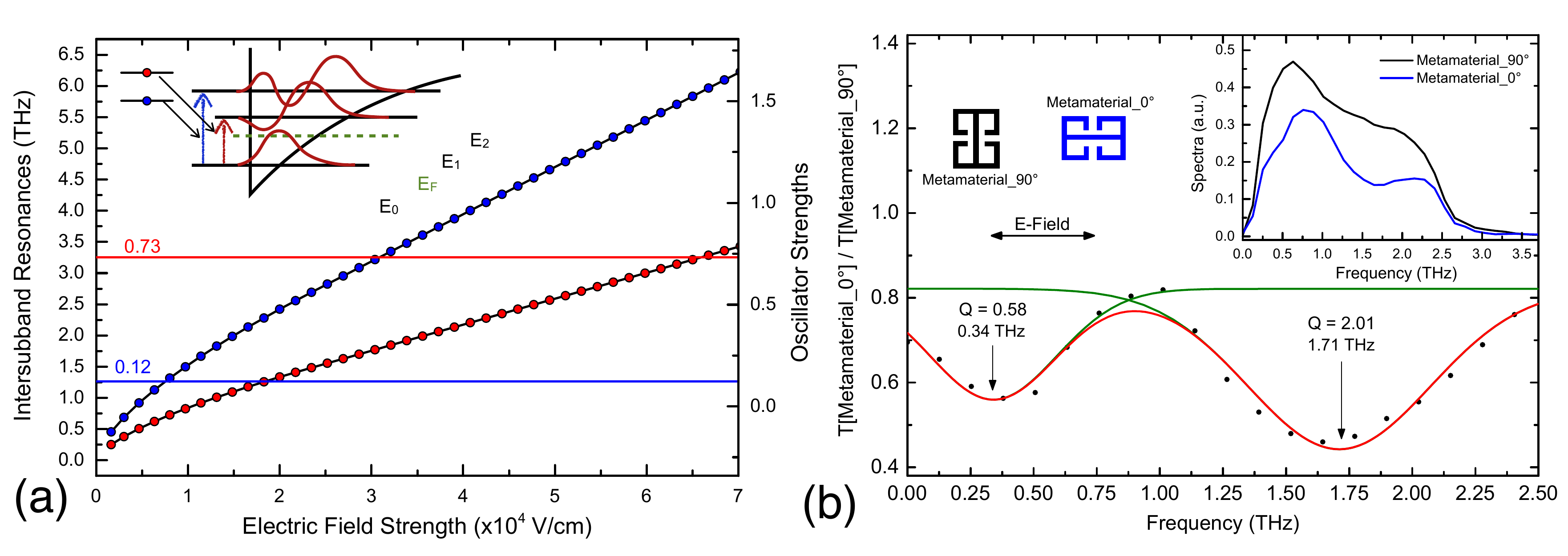}
\caption{{\bf{1D Poisson simulation and characterization of metamaterials.}} \textbf{(a)} Dependence of the intersubband resonances and the oscillator strengths for two intersubband transitions, $\left| 0 \right\rangle  \to \left| 1 \right\rangle$ and $\left| 0 \right\rangle  \to \left| 2 \right\rangle$, on the electric field strength and hence the bias applied on the gate. The oscillator strengths do not vary with the gate voltage and are fixed for each transition, which is a characteristic feature of the traingular potential well. Inset: Band-schematic of the first three subbands in a 2DEG with respect to the Fermi level at an intermediate gate voltage where only the ground subband is below the Fermi level. \textbf{(b)} The transmission of the incident polarized electric field for the horizonzal orientation of the metamaterial (Metamaterial$\char`_0^{\circ}$) normalized to the transmission for the vertical orientation of the metamaterial (Metamaterial$\char`_90^{\circ}$). The green spectra are results of deconvolution and the red curve is the reconstructed transmission spectrum. Inset: Spectra corresponding to the two orientations of the metamaterials. The schematic shows the orientations of the meta-structures with respect to the electric field direction in the THz TDS measurements.}
\label{fig:tds}
\end{figure*}

\noindent{{\bf 2DEG Intersubband Resonances}}\\
In order to characterize the energy eigenvalues of the 2DEG confined in a triangular potential well, a self-consistent Schr\"{o}dinger-Poisson equation is solved using the 1D Poisson solver \cite{Snider1990jap}. The solution is carried out for different bias applied to the gate. The dependence of the intersubband spacings of two resonances, $E_0 \to E_1$ and $E_0 \to E_2$ are plotted as a function of the electric field strength in fig.~\ref{fig:tds}(a), where $E_0$, $E_1$ and $E_2$ corresponds to the ground, first and second subbands respectively. By tuning the gate voltage, the confinement potential and hence the electric field strength, $F$ can be altered. Under the triangular well approximation \cite{Ando1976prb}, the shift of the ISR to higher energies is given by \cite{Wieck1989prb}:
\begin{equation}
{E_i} = {\left[ {\frac{{9{\pi ^2}}}{{8{m^*}}}} \right]^{\frac{1}{3}}}{\left( {e\hbar {F}} \right)^{\frac{2}{3}}}{\left( {i + \frac{3}{4}} \right)^{\frac{2}{3}}},
\end{equation}
where $i$ (0, 1, 2, ...) represent the indices of the subbands, $m^*$ is the effective mass of electrons in GaAs and $e$ is the electronic charge. The normalized oscillator strengths, $f_{ij}$, are calculated from the transition matrix elements, ${\left\langle j \right|z\left| i \right\rangle }$, corresponding to the ISRs by using the following relation:
\begin{equation}
{f_{ij}} = \frac{{2{m^*}{\omega _{ij}}}}{\hbar }{\left| {\left\langle j \right|z\left| i \right\rangle } \right|^2},
\end{equation}
where ${\left| {\left\langle j \right|z\left| i \right\rangle } \right|}$ is given by \cite{Wieck1989prb}:
\begin{equation}
\left| {\left\langle j \right|z\left| i \right\rangle } \right| = {z_{ji}} = \frac{{2L}}{{{{\left( {{t_i} - {t_{j}}} \right)}^2}}},
\end{equation}
with ${t_i} =  - {\left[ {{{3\pi \left( {i + \frac{3}{4}} \right)} \mathord{\left/
				{\vphantom {{3\pi \left( {i + \frac{3}{4}} \right)} 2}} \right.
				\kern-\nulldelimiterspace} 2}} \right]^{{\raise0.7ex\hbox{$2$} \!\mathord{\left/
				{\vphantom {2 3}}\right.\kern-\nulldelimiterspace}
			\!\lower0.7ex\hbox{$3$}}}}$. 
The quantity, $L$ is defined as the electric length, given by $L = {\left( {{{{\hbar ^2}} \mathord{\left/{\vphantom {{{\hbar ^2}} {2{m^*}eF}}} \right.\kern-\nulldelimiterspace} {2{m^*}eF}}} \right)^{1/3}}$. The normalized oscillator strength for $E_0 \to E_1$ transition is calculated to be 0.73 while that of $E_0 \to E_2$ transition is 0.12. According to the sum rule, the normalized oscillator strengths of all transitions sum up to 1, which indicates transition to higher levels are very weak. Under the framework of triangular well approximation, the oscillator strengths is calculated to be independent of the electric field strength (fig.~\ref{fig:tds}(a)). This is unique compared to that of the square potential well as shown by Benz et al.~\cite{Benz2013apl}, where there is a small contribution of the quantum-confined Stark shift to the oscillator strengths in the presence of external bias. Due to asymmetric confinement, the ISRs can be tuned with a magnitude one order higher as compared to an equivalent square well. For a triangular well with an effective thickness of 20 nm, if the electric field is tuned by 4.5 $\times$ 10$^4$ V/cm, the ISR ($E_0 \to E_1$) can be tuned by 35 meV, while for an equivalent square well for the same change in electric field the ISR is tuned only by 3 meV. The experimentally observed ISRs are blue-shifted due to resonant screening from higher subbands, also known as the depolarization effect \cite{Ando1976prb,Wieck1988prb,Wieck1989prb} and is explained in the methods section.\\

\noindent{{\bf Results}\\
\noindent{{\bf Characterization of metamaterials}}\\
\begin{figure}[b!]
\includegraphics[width=\columnwidth]{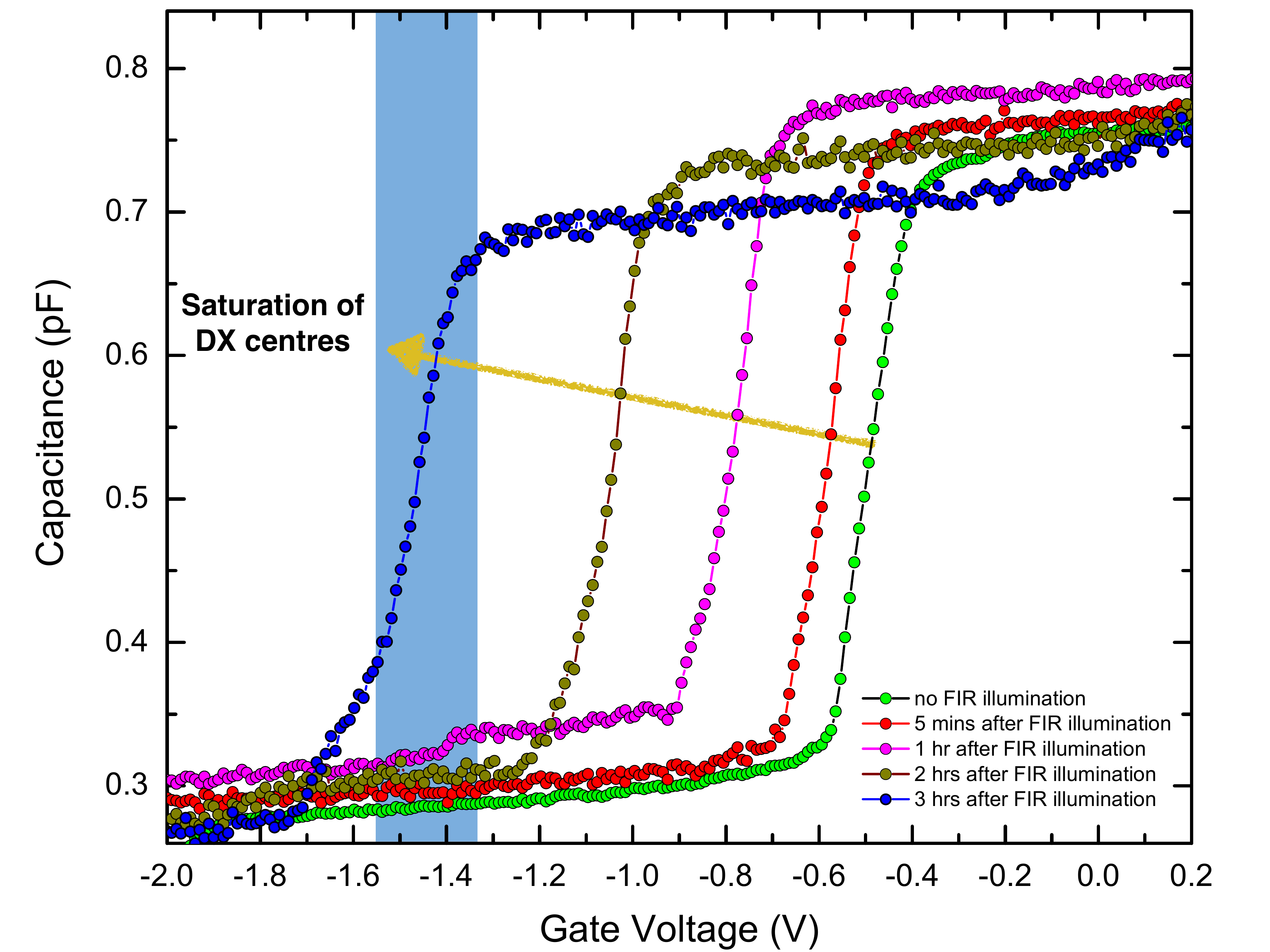}
\caption{{\bf{Capacitance-Voltage spectroscopy.}} Dependence of capacitance on the voltage applied to the gate. With illumination, visible part of the source saturates the DX centres and hence the threshold voltage shifts to more negative values. When a quasi-stable condition is reached, and the thereshold does not shift any further, all donors are saturated. The blue shaded region highlights the range of voltage over which the FTIR transmission spectra are taken.}
\label{fig:cap}
\end{figure}
\begin{figure*}[t!]
\includegraphics[width=\linewidth]{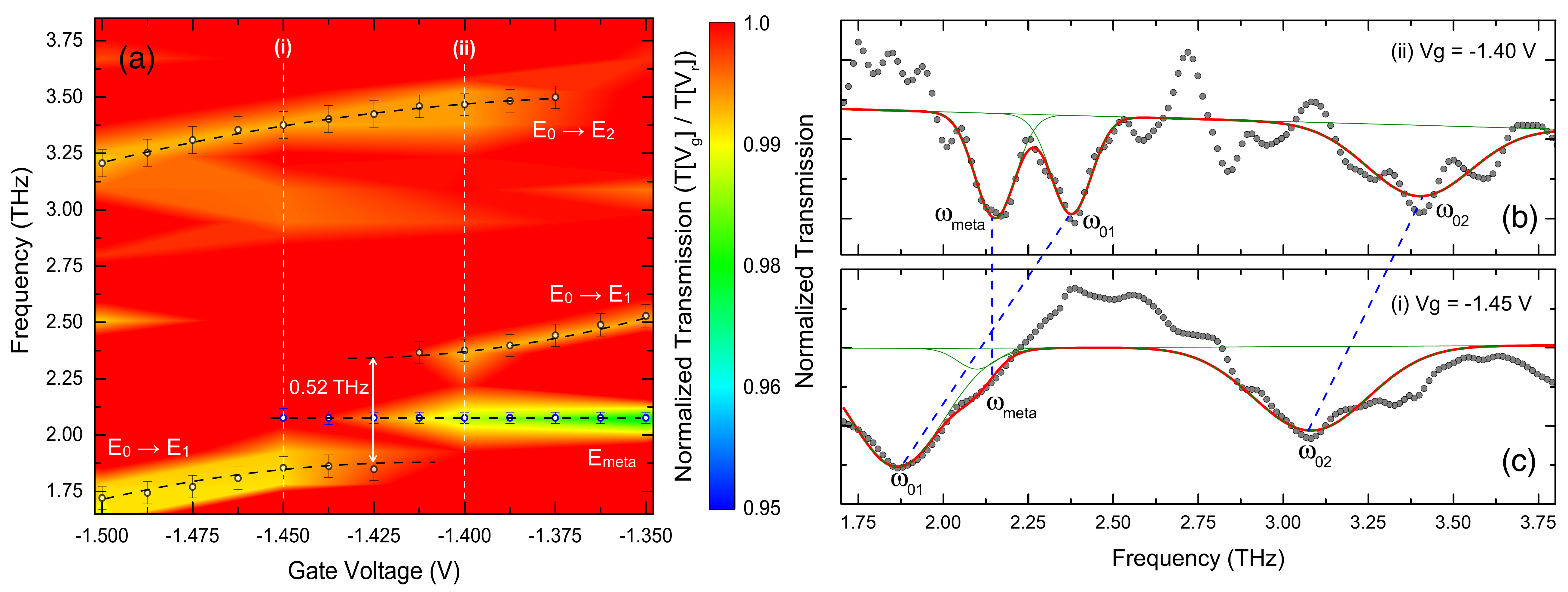}
\caption{{\bf{Observation of strong coupling.}} \textbf{(a)} Strong coupling is observed in the transmission spectra by tuning the gate voltages for the $\left| 0 \right\rangle  \to \left| 1 \right\rangle$ transition with the metamaterial resonance at around 2 THz, while no coupling is observed for the other intersubband transition, $\left| 0 \right\rangle  \to \left| 2 \right\rangle$. The coupling of the cavity resonance with the 2DEG intersubband resonance is seen as a split at the cross-over point between the two resonances with a split-gap of 0.52 THz. The normalized transmisison spectra at two voltage points, (i) $V_g = -1.45$ V and (ii) $V_g = -1.40$ V are shown in \textbf{(b)} and \textbf{(c)} respectively. Each spectral deconvolution shows three dips ($\omega_{01}$, $\omega_{meta}$ and $\omega_{02}$) corresponding to the points shown in (a). The blue dashed lines show the shift of the respective transitions.}
\label{fig:ftir}
\end{figure*}
The MMs are characterized by THz time-domain spectroscopy (TDS) at room temperature under purged conditions, to avoid the water absorption lines. The experimental details of the set-up are explained seperately in the methods section. At room temperature the thermal energy, $k_BT$ (= 25 meV), is greater than the subband energy spacings. This results in the occupation of higher subbands, which consequently prevents us from observing the ISR ($E_0 \to E_1$) from the 2DEG. Hence under this condition, the response from the sample is purely due to MMs. The transmission is measured for two orientations of the MMs, fig.~\ref{fig:tds}(b). In the inset, the spectra corresponding to the two orientations of MMs are shown. The electric field of the THz radiation couples to the MMs only when they are oriented parallel to the optical table, marked as Metamaterial$\char`_0^{\circ}$ in fig.~\ref{fig:tds}(b). The normalized transmission is plotted by dividing the transmitted intensity with respect to the MM orientation when the fields do not couple. Two strong resonances are observed in the transmission spectrum at 0.34 THz and 1.71 THz. The quality factors of the resonance frequencies at 0.34 THz and 1.71 THz are found to be 0.58 and 2.01 respectively. These resonances have different origin described as follows: The lower frequency at 0.34 THz is due to the LC equivalent circuit where the current circulates in the inductive part and the field is concentrated between the capacitor plates. The resonance at 1.71 THz is due to the half wave resonance \cite{Padilla2006prl} arising from the sides of MM. There are additional losses arising from the epitaxial layer, damping and change in the effective dielectric constant of the layers with the change in temperature. From the quality factor of the resonance at 1.71 THz, we estimated the losses to be around 0.85 THz. On cooling the sample, the cavity resonance will be blue shifted according to \cite{Ginn2009jap}: ${\omega _0} = \sqrt {\omega _r^2 + {\gamma ^2}}$, where $\omega_0$ is the bare cavity resonance without any losses, $\omega_r$ is the resonance of the MMs including losses and $\gamma$ is proportional to the ohmic losses in the metal and radiative damping in the cavity. The resulting resonance frequency without losses is calculated to be 1.9 THz, which is very close to the observed resonance at 4.2 K.\\

\noindent{{\bf Charging spectroscopy of the 2DEG}}\\
The charging of the 2DEG subbands is observed by capacitance-voltage spectroscopy at 4.2 K, shown in fig.~\ref{fig:cap}. The change in capacitance between the top gate and the bottom 2DEG layer is measured as a function of the gate voltage by applying DC + AC voltage to the gate and measuring the AC component of the current across the ohmics. The steep increase of capacitance in the voltage ranging from $-0.6$ V to $-0.4$ V indicates that the 2DEG layer is filled with charge carriers (green curve). When the sample is illuminated with far-infrared (FIR) broad-band source (Hg-arc lamp), the threshold voltage shifts more to negative values. This is due to the activation of the donor-exchange (DX) centres \cite{Mooney1990jap}. As the sample is illuminated longer, more donor atoms from the $\delta$-doped AlGaAs layer ionize and less bias is required to charge the 2DEG subbands, resulting in the shift of charging slope to more negative biases. When the charging spectrum does not shift any more (shown by the blue curve), a steady state is reached. The activation of the DX-centres is a result of residual near band-gap illumination (after filtering by black polyethylene window) from the FIR source. The blue shaded region in fig.~\ref{fig:cap} marks the voltage range where density-chopped infrared transmission measurements are performed. The Schr\"{o}dinger-Poisson equations are solved over the shaded voltage range, corresponding to the electric field strengths plotted in fig.~\ref{fig:tds}(a).\\

\noindent{{\bf Strong coupling of metamaterials with 2DEG}}\\
An FTIR transmission set-up with a rapid scan BRUKER IFS113V interferometer is used for the density chopped infrared transmission spectroscopy to observe the strong coupling of the MM cavity resonance with the 2DEG ISRs. As observed in the charging spectrum, at a bias of $-2$ V, the conduction band of the 2DEG is pulled above the Fermi level and hence completely depleted of carriers. This voltage is used as the reference voltage in the chopping-scheme at which sample transmission is recorded. The voltage is then increased to $-1.50$ V, where only one subband is below the Fermi level and corresponding transmission measurements are performed. The voltages are changed alternatively and the respective spectra are co-added and averaged over time. This is repeated for several gate voltages, keeping the reference voltage same. All the spectra are collectively plotted in a contour-plot in fig.~\ref{fig:ftir}(a). It is observed that at low temperature, the cavity resonance shifted to $2.12$ THz due to lower losses as compared to the room temperature measurements. The Q-factor of the MM resonance has dramatically improved to a value of 16.9. At $V_g = -1.425$ V, a clear splitting of the ISR can be observed when the ISR ($\omega_{01}$) of the 2DEG crosses the resonance of the MMs at 2.12 THz. The width of the splitting is found to be $0.52$ THz. Two normalized transmission spectra at $V_g = -1.45$ V and $V_g = -1.40$ V, shown by the white line in fig.~\ref{fig:ftir}(a) are plotted in fig.~\ref{fig:ftir}(b) and (c). The spectra are smoothed in order to carry out the spectral deconvolution. Three peaks, marked as $\omega_{01}$, $\omega_{meta}$ and $\omega_{02}$ can be seen corresponding to the points marked in fig.~\ref{fig:ftir}(a). As described before, the ISRs are blue shifted by $1.2$ meV due to the resonant screening from the higher subbands. It is observed that at $V_g = -2$ V, not only the 2DEG channel gets completely depleted of charge carriers but also the response from the capacitive element of the MMs is screened due to charges in the epitaxial layer. By applying bias on the epitaxial gate below the MM, a significant modulation of the intensity (or amplitude) of the cavity resonance is observed \cite{Shrekenhamer2011oe,Chen2006nl}. Thus in our chopping scheme, the MM resonance at $2.12$ THz disappears for biases lower than $-1.45$ V and begins to appear for biases higher than $-1.425$ V. In our experiments, we limit the bias applied on the gate to $-1.3$ V. Since with increasing bias the slope of the triangular confinement becomes more steeper, the second subband is pulled below the Fermi level, which makes the transition scheme different, which is not desirable.\\

\noindent{{\bf Discussion}}\\
The 2DEG and MMs in our device can be considered as two oscillators, one of which has a fixed frequency and the frequency of the other oscillator (2DEG) is tuned by the gate voltage. When the frequencies of both the oscillators are similar, they form a coupled system and an anti-crossing phenomenon is observed. Using the common oscillator model, described by Gabbay et al.~\cite{Gabbay2012oe}, the strength of this coupling is found to be directly proportional to the strength of the splitting, described in the methods section. The quantity, $\Omega$, defined as the coupling strength is given by:
\begin{equation}
\Omega  = {\omega _ + } - {\omega _ - } = 2\sqrt {\frac{1}{4}\Omega _0^2 - {{\left[ {\left( {{\xi _{meta}} - {\xi _{10}}} \right)} \right]}^2}},
\end{equation}
where $\Omega_0$ is the bare coupling strength, $\xi _{meta}$ and $\xi _{10}$ represent the losses in the metamterials and ISRs and can be equated to the linewidths of the respective transitions. The coupling strength in our experiment is found to be $0.52$ THz, while the bare coupling strength is calculated to be $0.56$ THz. There is no significant change in the two coupling strengths. The ratio of the splitting to the sum of full-width at half maximum of both the resonances ($\frac{\Omega }{{\Delta {\omega _{meta}} + \Delta {\omega _{10}}}} = \frac{{0.52}}{{0.36}} = 1.44$) is found to be greater than one. The normalized coupling ratio ($\frac{\Omega }{{{\omega _0}}}$) in our experiment is found to be $0.26$, which is a clear indication that the observed coupling is in the strong regime and is a significantly high value achieved for a single well.

In conclusion, we show strong coupling of the 2DEG ISR in a heterostructure with a triangular confining potential to the cavity resonance of a THz MM by driving the device from an uncoupled state to a strongly coupled state via electrical tuning of ISR. Due to asymmetric confinement potential, the tuning of the ISR is found to be one order of magnitude higher in comparison to an identical square well. The use of epitaxial gate proved to be advantageous for the use of the direct MMs and consequently tune the ISR at the same time. With a proper chopping scheme, we successfully demonstrated that the measurement can be performed in one integrated sample without the need of additional reference sample. This is one of the remarkable feature of our device. Moreover, we succeded to achieve ultrastrong light-matter interaction by employing a single triangular quantum well in a high electron mobility transistor heterostructure, with a normalized coupling ratio of 0.26.\\

\noindent{{\bf Methods}}\\
\noindent{{\bf Sample Design}}\\
The sample is grown on a semi-insulating GaAs (100) substrate by molecular beam epitaxy. At first 20 periods of AlAs/GaAs (5 nm/5 nm) short period superlattice (SPS) are grown to smoothen the surface for growth. A 650 nm thick GaAs layer is grown followed by a 50 nm Al$_{0.34}$Ga$_{0.66}$As spacer layer. This is followed by a Si-$\delta$ doping and then a 45 nm Al$_{0.34}$Ga$_{0.66}$As layer. Another 6 periods of SPS layer comprising of 1 nm GaAs and 3 nm AlAs are grown, capped by a 13 nm GaAs layer above which the epitaxial gate \cite{Pal2014jpcm} is grown. The epitaxial gate has a 15 nm thick carbon-doped GaAs layer and 30 periods of carbon-$\delta$ doped and 0.5 nm carbon-doped GaAs layers, which lead to an effective carrier density of N$_A$ = 2 $\times$ 10$^{-19}$ cm$^{-3}$. To establish a contact with the 2DEG layer, 162 nm below the surface, the corners of a 5 mm $\times$ 6 mm sample are first etched down by 100 nm (fig.~\ref{fig:device}(b)) and then indium is diffused in a reaction chamber in an inert atmosphere of argon and nitrogen. The sample is then mounted on a chip carrier and wires are bonded by wedge-bonding.  The structures are processed on the epitaxial gate, shown schematically in fig.~\ref{fig:device}(b), by conventional UV-photolithography, evaporation of Cr/Au (10/200 nm) and lift-off technique. Scanning electron micrographs (SEM) of the MMs and a wire bonding are shown in fig.~\ref{fig:device}(c) and (d).\\

\noindent{{\bf THz time domain spectroscopy}}\\
A Ti:Sa laser with 80 fs pulse duration (centre wavelength of 800 nm) and a repetition rate of 80 MHz is used to generate THz radiation by exciting an inter-digitated photoconductive antenna processed on a GaAs substrate at an applied DC bias. The experiment is performed in a transmission geometry with an NIR-power of 200 mW on the antenna. The detection is based on electro-optic sampling of the THz electric field by employing a birefringent, 2 mm thick ZnTe crystal. The electric field component of the source is in a direction parallel to the optical table. Two 90$^\circ$ off-axis parabolic mirrors are used to collimate and focus the THz beam on the sample, which is fixed on a rotation mount. Another two parabolic mirrors are used to collect the signal from the sample and focus it on the ZnTe crystal. The transmission measurements are performed to observe the response of the metamaterials alone in the device under purged conditions.\\

\noindent{{\bf Oscillator Strengths and Depolarization Shifts}}\\
Under the triangular well approximation, the intersubband energy edges, $E_i$, are directly proportional to $F^{2/3}$, where $F$ is the electric field strength. Now the oscillator strength for the transition between two subbands in a triangular potential well is proportional to the square of the matrix elements, which in turn is proportional to the square of the electric length, $L$. The electric length is inversely proportional to $F^{1/3}$. This implies that the oscillator strength does not depend on the field strength in a triangular well and is a constant given by:
\begin{equation}
{f_{ij}} = {\left( {\frac{\pi }{2}} \right)^{2/3}}\frac{12}{{{{\left( {{t_i} - {t_j}} \right)}^4}}}\left[ {{{\left( {j + \frac{3}{4}} \right)}^{2/3}} - {{\left( {i + \frac{3}{4}} \right)}^{2/3}}} \right]
\end{equation}
where, ${t_i} =  - {\left[ {{{3\pi \left( {i + \frac{3}{4}} \right)} \mathord{\left/
				{\vphantom {{3\pi \left( {i + \frac{3}{4}} \right)} 2}} \right.
				\kern-\nulldelimiterspace} 2}} \right]^{2/3}}$ \cite{Wieck1989prb}. 
For $i = 0$ and $j=1$, $f_{01}$ = 0.73 and for $i = 0$ and $j=2$, $f_{02}$ = 0.12. Taking in to account the allowed transitions to higher subbands ($j = 3, 4, ...$), the total oscillator strength, $\sum\limits_j {{f_{0j}}}$, sums up to 1. The normalized energy eigenfunctions for the ground and the first excited intersubbands under triangular confinement, are given by \cite{Wieck1989prb}:
\begin{equation}
{\phi _i}\left( z \right) = \frac{1}{{\sqrt {{N_i}} }}Ai\left( {\frac{z}{L} - \frac{{{E_i}}}{{eFL}}} \right),
\end{equation}
where, $N_i$ are the normalization constants given by $Ai'\left( {{t_i}} \right)$ and $L = {\left( {{{{\hbar ^2}} \mathord{\left/{\vphantom {{{\hbar ^2}} {2{m^*}eF}}} \right.\kern-\nulldelimiterspace} {2{m^*}eF}}} \right)^{1/3}}$. Including the resonant screening effects, the intersubband energies are blue shifted by
\begin{equation}
{\tilde E_{10}} = {E_{10}}\sqrt {\left( {1 + {\alpha _{11}}} \right)},
\end{equation}
where $\alpha_{11}$ is the depolarization factor, which according to Ando~\cite{Ando1977ssc} is given by:
\begin{equation}
{\alpha _{11}} = \frac{{8\pi {e^2}{n_{2d}}}}{{\epsilon {\epsilon _0}}}{S_{11}}\frac{1}{E_{10}},
\end{equation}
where $n_{2d}$ is the 2D carrier density; $\epsilon$ is the dielectric constant of GaAs; $\epsilon_0$ is the permittivity of free space. $\hbar{\omega_{10}}$ is the energy spacing corresponding to the transition,  $E_0 \to E_1$ and the overlap factor, $S_{11}$ has the dimensions of length and is given by \cite{Ando1977ssc}:
\begin{equation}
{S_{11}} = \int\limits_0^\infty  {dz{{\left[ {\int\limits_0^z {dz'{\phi _1}\left( {z'} \right){\phi _0}\left( {z'} \right)} } \right]}^2}},
\end{equation}
which can be written as:
\begin{equation}
{S_{11}} = {\left( {\frac{\hbar^2}{{2{m^*}{E_{10}}}}} \right)^2}\int\limits_0^\infty {dz{{\left[ {{{\phi '}_1}\left( z \right){\phi _0}\left( z \right) - {\phi _1}\left( z \right){{\phi '}_0}\left( z \right)} \right]}^2}}.
\end{equation}
The carrier density is however not a fixed value and changes with the change in bias. Besides the bias also changes the slope of the confinement thus changing the overlap-integral ($S_{11}$). At a bias of $-1.4$ V, corresponding to a carrier density of 2 $\times$ 10$^{11}$ cm$^{-2}$, the depolarization shift is calculated to be 1.2 meV. The details of the calculation is shown in the methods section. The influence of resonant screening becomes important when the level spacings are smaller and the carrier densities are high.\\

\noindent{{\bf Coupling Strength}}\\
The system is assumed to be comprised of two oscillators which are tuned from uncoupled to strongly coupled regime by tuning the ISR of the 2DEG with an external gate bias, making use of quantum-confined Stark effect. Following the model of coupled oscillators, as developed by Gabbay et al.~\cite{Gabbay2012oe}, the coupling between the two resonators can be described by a simple 2 $\times$ 2 matrix:
\begin{equation}
\left( {\begin{array}{*{20}{c}}
	{{\omega _{meta}} - i{\xi _{meta}}}&{\Omega_0 /2}\\
	{\Omega_0 /2}&{{\omega _{10}} - i{\xi _{10}}}
	\end{array}} \right),
\end{equation}
where $\omega_{10}$, $\omega_{meta}$ are the resonances and $\xi_{10}$, $\xi_{meta}$ are the damping factors corresponding to ISR and metamaterials; $\Omega_0$ is the bare coupling strength of the oscillators. The matrix can be diagonalized to get the upper and lower eigenvalues of the coupled system as:
\begin{equation}
\begin{array}{l}
{\omega _ \pm } = \frac{{{\omega _{meta}} + {\omega _{10}} + i\left( {{\xi _{meta}} + {\xi _{10}}} \right)}}{2} \pm 
\sqrt {\frac{1}{4}{\Omega_0 ^2} + {{\left[ {{\omega _{meta}} - {\omega _{10}} + i\left( {{\xi _{meta}} - {\xi _{10}}} \right)} \right]}^2}},
\end{array}
\end{equation}
Subtracting the frequencies, we obtain:
\begin{equation}
{\omega _ + } - {\omega _ - } = 2\sqrt {\frac{1}{4}{\Omega_0 ^2} + {{\left[ {{\omega _{meta}} - {\omega _{10}} + i\left( {{\xi _{meta}} - {\xi _{10}}} \right)} \right]}^2}},
\end{equation}
At the anticrossing point, both the resonators start oscillating with equal and real frequencies, the coupling strength, $\Omega$, can be expressed as:
\begin{equation}
\Omega  = {\omega _ + } - {\omega _ - } = 2\sqrt {\frac{1}{4}\Omega _0^2 - {{\left[ {\left( {{\xi _{meta}} - {\xi _{10}}} \right)} \right]}^2}}.
\end{equation}
From the two (upper and lower) branches of the transitions, the coupling strength, $\Omega$, is found to be 0.52 THz. The damping factors corresponding to the resonances are $\xi_{meta} = 0.13$ THz and $\xi_{10} = 0.23$ THz. The bare coupling strength, $\Omega_0$, is calculated to be 0.56 THz which similar to the coupling strength indicating strong coupling with low losses.\\

\bibliography{meta}

\begin{thebibliography}{42}%
\makeatletter
\providecommand \@ifxundefined [1]{%
 \@ifx{#1\undefined}
}%
\providecommand \@ifnum [1]{%
 \ifnum #1\expandafter \@firstoftwo
 \else \expandafter \@secondoftwo
 \fi
}%
\providecommand \@ifx [1]{%
 \ifx #1\expandafter \@firstoftwo
 \else \expandafter \@secondoftwo
 \fi
}%
\providecommand \natexlab [1]{#1}%
\providecommand \enquote  [1]{``#1''}%
\providecommand \bibnamefont  [1]{#1}%
\providecommand \bibfnamefont [1]{#1}%
\providecommand \citenamefont [1]{#1}%
\providecommand \href@noop [0]{\@secondoftwo}%
\providecommand \href [0]{\begingroup \@sanitize@url \@href}%
\providecommand \@href[1]{\@@startlink{#1}\@@href}%
\providecommand \@@href[1]{\endgroup#1\@@endlink}%
\providecommand \@sanitize@url [0]{\catcode `\\12\catcode `\$12\catcode
  `\&12\catcode `\#12\catcode `\^12\catcode `\_12\catcode `\%12\relax}%
\providecommand \@@startlink[1]{}%
\providecommand \@@endlink[0]{}%
\providecommand \url  [0]{\begingroup\@sanitize@url \@url }%
\providecommand \@url [1]{\endgroup\@href {#1}{\urlprefix }}%
\providecommand \urlprefix  [0]{URL }%
\providecommand \Eprint [0]{\href }%
\providecommand \doibase [0]{http://dx.doi.org/}%
\providecommand \selectlanguage [0]{\@gobble}%
\providecommand \bibinfo  [0]{\@secondoftwo}%
\providecommand \bibfield  [0]{\@secondoftwo}%
\providecommand \translation [1]{[#1]}%
\providecommand \BibitemOpen [0]{}%
\providecommand \bibitemStop [0]{}%
\providecommand \bibitemNoStop [0]{.\EOS\space}%
\providecommand \EOS [0]{\spacefactor3000\relax}%
\providecommand \BibitemShut  [1]{\csname bibitem#1\endcsname}%
\let\auto@bib@innerbib\@empty
\bibitem [{\citenamefont {Wallraff}\ \emph {et~al.}(2004)\citenamefont
  {Wallraff}, \citenamefont {Schuster}, \citenamefont {Blais}, \citenamefont
  {Frunzio}, \citenamefont {Huang}, \citenamefont {Majer}, \citenamefont
  {Kumar}, \citenamefont {Girvin},\ and\ \citenamefont
  {Schoelkopf}}]{Wallraff2004n}%
  \BibitemOpen
  \bibfield  {author} {\bibinfo {author} {\bibfnamefont {A.}~\bibnamefont
  {Wallraff}}, \bibinfo {author} {\bibfnamefont {D.~I.}\ \bibnamefont
  {Schuster}}, \bibinfo {author} {\bibfnamefont {A.}~\bibnamefont {Blais}},
  \bibinfo {author} {\bibfnamefont {L.}~\bibnamefont {Frunzio}}, \bibinfo
  {author} {\bibfnamefont {R.-S.}\ \bibnamefont {Huang}}, \bibinfo {author}
  {\bibfnamefont {J.}~\bibnamefont {Majer}}, \bibinfo {author} {\bibfnamefont
  {S.}~\bibnamefont {Kumar}}, \bibinfo {author} {\bibfnamefont {S.~M.}\
  \bibnamefont {Girvin}}, \ and\ \bibinfo {author} {\bibfnamefont {R.~J.}\
  \bibnamefont {Schoelkopf}},\ }\href {\doibase 10.1038/nature02851} {\bibfield
   {journal} {\bibinfo  {journal} {Nature}\ }\textbf {\bibinfo {volume}
  {431}},\ \bibinfo {pages} {162} (\bibinfo {year} {2004})}\BibitemShut
  {NoStop}%
\bibitem [{\citenamefont {Macha}\ \emph {et~al.}(2014)\citenamefont {Macha},
  \citenamefont {Oelsner}, \citenamefont {Reiner}, \citenamefont {Marthaler},
  \citenamefont {André}, \citenamefont {Schön}, \citenamefont {Hübner},
  \citenamefont {Meyer}, \citenamefont {Il’ichev},\ and\ \citenamefont
  {Ustinov}}]{Macha2014nc}%
  \BibitemOpen
  \bibfield  {author} {\bibinfo {author} {\bibfnamefont {P.}~\bibnamefont
  {Macha}}, \bibinfo {author} {\bibfnamefont {G.}~\bibnamefont {Oelsner}},
  \bibinfo {author} {\bibfnamefont {J.-M.}\ \bibnamefont {Reiner}}, \bibinfo
  {author} {\bibfnamefont {M.}~\bibnamefont {Marthaler}}, \bibinfo {author}
  {\bibfnamefont {S.}~\bibnamefont {André}}, \bibinfo {author} {\bibfnamefont
  {G.}~\bibnamefont {Schön}}, \bibinfo {author} {\bibfnamefont
  {U.}~\bibnamefont {Hübner}}, \bibinfo {author} {\bibfnamefont {H.-G.}\
  \bibnamefont {Meyer}}, \bibinfo {author} {\bibfnamefont {E.}~\bibnamefont
  {Il’ichev}}, \ and\ \bibinfo {author} {\bibfnamefont {A.~V.}\ \bibnamefont
  {Ustinov}},\ }\href {\doibase 10.1038/ncomms6146} {\bibfield  {journal}
  {\bibinfo  {journal} {Nat Commun}\ }\textbf {\bibinfo {volume} {5}} (\bibinfo
  {year} {2014}),\ 10.1038/ncomms6146}\BibitemShut {NoStop}%
\bibitem [{\citenamefont {Ye}\ \emph {et~al.}(1999)\citenamefont {Ye},
  \citenamefont {Vernooy},\ and\ \citenamefont {Kimble}}]{Kimble1999prl}%
  \BibitemOpen
  \bibfield  {author} {\bibinfo {author} {\bibfnamefont {J.}~\bibnamefont
  {Ye}}, \bibinfo {author} {\bibfnamefont {D.~W.}\ \bibnamefont {Vernooy}}, \
  and\ \bibinfo {author} {\bibfnamefont {H.~J.}\ \bibnamefont {Kimble}},\
  }\href {\doibase 10.1103/PhysRevLett.83.4987} {\bibfield  {journal} {\bibinfo
   {journal} {Phys. Rev. Lett.}\ }\textbf {\bibinfo {volume} {83}},\ \bibinfo
  {pages} {4987} (\bibinfo {year} {1999})}\BibitemShut {NoStop}%
\bibitem [{\citenamefont {Kimble}(2008)}]{Kimble2008n}%
  \BibitemOpen
  \bibfield  {author} {\bibinfo {author} {\bibfnamefont {H.~J.}\ \bibnamefont
  {Kimble}},\ }\href {\doibase 10.1038/nature07127} {\bibfield  {journal}
  {\bibinfo  {journal} {Nature}\ }\textbf {\bibinfo {volume} {453}},\ \bibinfo
  {pages} {1023} (\bibinfo {year} {2008})}\BibitemShut {NoStop}%
\bibitem [{\citenamefont {McKeever}\ \emph {et~al.}(2003)\citenamefont
  {McKeever}, \citenamefont {Boca}, \citenamefont {Boozer}, \citenamefont
  {Buck},\ and\ \citenamefont {Kimble}}]{Mckeever2003n}%
  \BibitemOpen
  \bibfield  {author} {\bibinfo {author} {\bibfnamefont {J.}~\bibnamefont
  {McKeever}}, \bibinfo {author} {\bibfnamefont {A.}~\bibnamefont {Boca}},
  \bibinfo {author} {\bibfnamefont {A.~D.}\ \bibnamefont {Boozer}}, \bibinfo
  {author} {\bibfnamefont {J.~R.}\ \bibnamefont {Buck}}, \ and\ \bibinfo
  {author} {\bibfnamefont {H.~J.}\ \bibnamefont {Kimble}},\ }\href {\doibase
  10.1038/nature01974} {\bibfield  {journal} {\bibinfo  {journal} {Nature}\
  }\textbf {\bibinfo {volume} {425}},\ \bibinfo {pages} {268} (\bibinfo {year}
  {2003})},\ \bibinfo {note} {00451}\BibitemShut {NoStop}%
\bibitem [{\citenamefont {Hennessy}\ \emph {et~al.}(2007)\citenamefont
  {Hennessy}, \citenamefont {Badolato}, \citenamefont {Winger}, \citenamefont
  {Gerace}, \citenamefont {Atatüre}, \citenamefont {Gulde}, \citenamefont
  {Fält}, \citenamefont {Hu},\ and\ \citenamefont
  {Imamoğlu}}]{Hennessy2007n}%
  \BibitemOpen
  \bibfield  {author} {\bibinfo {author} {\bibfnamefont {K.}~\bibnamefont
  {Hennessy}}, \bibinfo {author} {\bibfnamefont {A.}~\bibnamefont {Badolato}},
  \bibinfo {author} {\bibfnamefont {M.}~\bibnamefont {Winger}}, \bibinfo
  {author} {\bibfnamefont {D.}~\bibnamefont {Gerace}}, \bibinfo {author}
  {\bibfnamefont {M.}~\bibnamefont {Atatüre}}, \bibinfo {author}
  {\bibfnamefont {S.}~\bibnamefont {Gulde}}, \bibinfo {author} {\bibfnamefont
  {S.}~\bibnamefont {Fält}}, \bibinfo {author} {\bibfnamefont {E.~L.}\
  \bibnamefont {Hu}}, \ and\ \bibinfo {author} {\bibfnamefont {A.}~\bibnamefont
  {Imamoğlu}},\ }\href {\doibase 10.1038/nature05586} {\bibfield  {journal}
  {\bibinfo  {journal} {Nature}\ }\textbf {\bibinfo {volume} {445}},\ \bibinfo
  {pages} {896} (\bibinfo {year} {2007})}\BibitemShut {NoStop}%
\bibitem [{\citenamefont {Dini}\ \emph {et~al.}(2003)\citenamefont {Dini},
  \citenamefont {Köhler}, \citenamefont {Tredicucci}, \citenamefont
  {Biasiol},\ and\ \citenamefont {Sorba}}]{Dini2003prl}%
  \BibitemOpen
  \bibfield  {author} {\bibinfo {author} {\bibfnamefont {D.}~\bibnamefont
  {Dini}}, \bibinfo {author} {\bibfnamefont {R.}~\bibnamefont {Köhler}},
  \bibinfo {author} {\bibfnamefont {A.}~\bibnamefont {Tredicucci}}, \bibinfo
  {author} {\bibfnamefont {G.}~\bibnamefont {Biasiol}}, \ and\ \bibinfo
  {author} {\bibfnamefont {L.}~\bibnamefont {Sorba}},\ }\href {\doibase
  10.1103/PhysRevLett.90.116401} {\bibfield  {journal} {\bibinfo  {journal}
  {Phys. Rev. Lett.}\ }\textbf {\bibinfo {volume} {90}},\ \bibinfo {pages}
  {116401} (\bibinfo {year} {2003})}\BibitemShut {NoStop}%
\bibitem [{\citenamefont {Todorov}\ \emph {et~al.}(2009)\citenamefont
  {Todorov}, \citenamefont {Andrews}, \citenamefont {Sagnes}, \citenamefont
  {Colombelli}, \citenamefont {Klang}, \citenamefont {Strasser},\ and\
  \citenamefont {Sirtori}}]{Todorov2009prl}%
  \BibitemOpen
  \bibfield  {author} {\bibinfo {author} {\bibfnamefont {Y.}~\bibnamefont
  {Todorov}}, \bibinfo {author} {\bibfnamefont {A.~M.}\ \bibnamefont
  {Andrews}}, \bibinfo {author} {\bibfnamefont {I.}~\bibnamefont {Sagnes}},
  \bibinfo {author} {\bibfnamefont {R.}~\bibnamefont {Colombelli}}, \bibinfo
  {author} {\bibfnamefont {P.}~\bibnamefont {Klang}}, \bibinfo {author}
  {\bibfnamefont {G.}~\bibnamefont {Strasser}}, \ and\ \bibinfo {author}
  {\bibfnamefont {C.}~\bibnamefont {Sirtori}},\ }\href {\doibase
  10.1103/PhysRevLett.102.186402} {\bibfield  {journal} {\bibinfo  {journal}
  {Physical Review Letters}\ }\textbf {\bibinfo {volume} {102}} (\bibinfo
  {year} {2009}),\ 10.1103/PhysRevLett.102.186402}\BibitemShut {NoStop}%
\bibitem [{\citenamefont {Todorov}\ \emph {et~al.}(2010)\citenamefont
  {Todorov}, \citenamefont {Andrews}, \citenamefont {Colombelli}, \citenamefont
  {De~Liberato}, \citenamefont {Ciuti}, \citenamefont {Klang}, \citenamefont
  {Strasser},\ and\ \citenamefont {Sirtori}}]{Todorov2010prl}%
  \BibitemOpen
  \bibfield  {author} {\bibinfo {author} {\bibfnamefont {Y.}~\bibnamefont
  {Todorov}}, \bibinfo {author} {\bibfnamefont {A.~M.}\ \bibnamefont
  {Andrews}}, \bibinfo {author} {\bibfnamefont {R.}~\bibnamefont {Colombelli}},
  \bibinfo {author} {\bibfnamefont {S.}~\bibnamefont {De~Liberato}}, \bibinfo
  {author} {\bibfnamefont {C.}~\bibnamefont {Ciuti}}, \bibinfo {author}
  {\bibfnamefont {P.}~\bibnamefont {Klang}}, \bibinfo {author} {\bibfnamefont
  {G.}~\bibnamefont {Strasser}}, \ and\ \bibinfo {author} {\bibfnamefont
  {C.}~\bibnamefont {Sirtori}},\ }\href {\doibase
  10.1103/PhysRevLett.105.196402} {\bibfield  {journal} {\bibinfo  {journal}
  {Physical Review Letters}\ }\textbf {\bibinfo {volume} {105}} (\bibinfo
  {year} {2010}),\ 10.1103/PhysRevLett.105.196402}\BibitemShut {NoStop}%
\bibitem [{\citenamefont {Agarwal}(1984)}]{Agarwal1984prl}%
  \BibitemOpen
  \bibfield  {author} {\bibinfo {author} {\bibfnamefont {G.~S.}\ \bibnamefont
  {Agarwal}},\ }\href {\doibase 10.1103/PhysRevLett.53.1732} {\bibfield
  {journal} {\bibinfo  {journal} {Phys. Rev. Lett.}\ }\textbf {\bibinfo
  {volume} {53}},\ \bibinfo {pages} {1732} (\bibinfo {year}
  {1984})}\BibitemShut {NoStop}%
\bibitem [{\citenamefont {Agarwal}(1985)}]{Agarwal1985josab}%
  \BibitemOpen
  \bibfield  {author} {\bibinfo {author} {\bibfnamefont {G.~S.}\ \bibnamefont
  {Agarwal}},\ }\href {\doibase 10.1364/JOSAB.2.000480} {\bibfield  {journal}
  {\bibinfo  {journal} {J. Opt. Soc. Am. B}\ }\textbf {\bibinfo {volume} {2}},\
  \bibinfo {pages} {480} (\bibinfo {year} {1985})}\BibitemShut {NoStop}%
\bibitem [{\citenamefont {Zhu}\ \emph {et~al.}(1990)\citenamefont {Zhu},
  \citenamefont {Gauthier}, \citenamefont {Morin}, \citenamefont {Wu},
  \citenamefont {Carmichael},\ and\ \citenamefont {Mossberg}}]{Zhu1990prl}%
  \BibitemOpen
  \bibfield  {author} {\bibinfo {author} {\bibfnamefont {Y.}~\bibnamefont
  {Zhu}}, \bibinfo {author} {\bibfnamefont {D.~J.}\ \bibnamefont {Gauthier}},
  \bibinfo {author} {\bibfnamefont {S.~E.}\ \bibnamefont {Morin}}, \bibinfo
  {author} {\bibfnamefont {Q.}~\bibnamefont {Wu}}, \bibinfo {author}
  {\bibfnamefont {H.~J.}\ \bibnamefont {Carmichael}}, \ and\ \bibinfo {author}
  {\bibfnamefont {T.~W.}\ \bibnamefont {Mossberg}},\ }\href {\doibase
  10.1103/PhysRevLett.64.2499} {\bibfield  {journal} {\bibinfo  {journal}
  {Phys. Rev. Lett.}\ }\textbf {\bibinfo {volume} {64}},\ \bibinfo {pages}
  {2499} (\bibinfo {year} {1990})}\BibitemShut {NoStop}%
\bibitem [{\citenamefont {Meyrath}\ \emph {et~al.}(2007)\citenamefont
  {Meyrath}, \citenamefont {Zentgraf},\ and\ \citenamefont
  {Giessen}}]{Meyrath2007prb}%
  \BibitemOpen
  \bibfield  {author} {\bibinfo {author} {\bibfnamefont {T.~P.}\ \bibnamefont
  {Meyrath}}, \bibinfo {author} {\bibfnamefont {T.}~\bibnamefont {Zentgraf}}, \
  and\ \bibinfo {author} {\bibfnamefont {H.}~\bibnamefont {Giessen}},\ }\href
  {\doibase 10.1103/PhysRevB.75.205102} {\bibfield  {journal} {\bibinfo
  {journal} {Physical Review B}\ }\textbf {\bibinfo {volume} {75}} (\bibinfo
  {year} {2007}),\ 10.1103/PhysRevB.75.205102}\BibitemShut {NoStop}%
\bibitem [{\citenamefont {Scalari}\ \emph {et~al.}(2012)\citenamefont
  {Scalari}, \citenamefont {Maissen}, \citenamefont {Turcinkova}, \citenamefont
  {Hagenmuller}, \citenamefont {De~Liberato}, \citenamefont {Ciuti},
  \citenamefont {Reichl}, \citenamefont {Schuh}, \citenamefont {Wegscheider},
  \citenamefont {Beck},\ and\ \citenamefont {Faist}}]{Scalari2012sc}%
  \BibitemOpen
  \bibfield  {author} {\bibinfo {author} {\bibfnamefont {G.}~\bibnamefont
  {Scalari}}, \bibinfo {author} {\bibfnamefont {C.}~\bibnamefont {Maissen}},
  \bibinfo {author} {\bibfnamefont {D.}~\bibnamefont {Turcinkova}}, \bibinfo
  {author} {\bibfnamefont {D.}~\bibnamefont {Hagenmuller}}, \bibinfo {author}
  {\bibfnamefont {S.}~\bibnamefont {De~Liberato}}, \bibinfo {author}
  {\bibfnamefont {C.}~\bibnamefont {Ciuti}}, \bibinfo {author} {\bibfnamefont
  {C.}~\bibnamefont {Reichl}}, \bibinfo {author} {\bibfnamefont
  {D.}~\bibnamefont {Schuh}}, \bibinfo {author} {\bibfnamefont
  {W.}~\bibnamefont {Wegscheider}}, \bibinfo {author} {\bibfnamefont
  {M.}~\bibnamefont {Beck}}, \ and\ \bibinfo {author} {\bibfnamefont
  {J.}~\bibnamefont {Faist}},\ }\href {\doibase 10.1126/science.1216022}
  {\bibfield  {journal} {\bibinfo  {journal} {Science}\ }\textbf {\bibinfo
  {volume} {335}},\ \bibinfo {pages} {1323} (\bibinfo {year}
  {2012})}\BibitemShut {NoStop}%
\bibitem [{\citenamefont {Scalari}\ \emph {et~al.}(2013)\citenamefont
  {Scalari}, \citenamefont {Maissen}, \citenamefont {Hagenmüller},
  \citenamefont {De~Liberato}, \citenamefont {Ciuti}, \citenamefont {Reichl},
  \citenamefont {Wegscheider}, \citenamefont {Schuh}, \citenamefont {Beck},\
  and\ \citenamefont {Faist}}]{Scalari2013jap}%
  \BibitemOpen
  \bibfield  {author} {\bibinfo {author} {\bibfnamefont {G.}~\bibnamefont
  {Scalari}}, \bibinfo {author} {\bibfnamefont {C.}~\bibnamefont {Maissen}},
  \bibinfo {author} {\bibfnamefont {D.}~\bibnamefont {Hagenmüller}}, \bibinfo
  {author} {\bibfnamefont {S.}~\bibnamefont {De~Liberato}}, \bibinfo {author}
  {\bibfnamefont {C.}~\bibnamefont {Ciuti}}, \bibinfo {author} {\bibfnamefont
  {C.}~\bibnamefont {Reichl}}, \bibinfo {author} {\bibfnamefont
  {W.}~\bibnamefont {Wegscheider}}, \bibinfo {author} {\bibfnamefont
  {D.}~\bibnamefont {Schuh}}, \bibinfo {author} {\bibfnamefont
  {M.}~\bibnamefont {Beck}}, \ and\ \bibinfo {author} {\bibfnamefont
  {J.}~\bibnamefont {Faist}},\ }\href {\doibase 10.1063/1.4795543} {\bibfield
  {journal} {\bibinfo  {journal} {Journal of Applied Physics}\ }\textbf
  {\bibinfo {volume} {113}},\ \bibinfo {pages} {136510} (\bibinfo {year}
  {2013})}\BibitemShut {NoStop}%
\bibitem [{\citenamefont {Maissen}\ \emph {et~al.}(2014)\citenamefont
  {Maissen}, \citenamefont {Scalari}, \citenamefont {Valmorra}, \citenamefont
  {Beck}, \citenamefont {Faist}, \citenamefont {Cibella}, \citenamefont
  {Leoni}, \citenamefont {Reichl}, \citenamefont {Charpentier},\ and\
  \citenamefont {Wegscheider}}]{Maissen2014prb}%
  \BibitemOpen
  \bibfield  {author} {\bibinfo {author} {\bibfnamefont {C.}~\bibnamefont
  {Maissen}}, \bibinfo {author} {\bibfnamefont {G.}~\bibnamefont {Scalari}},
  \bibinfo {author} {\bibfnamefont {F.}~\bibnamefont {Valmorra}}, \bibinfo
  {author} {\bibfnamefont {M.}~\bibnamefont {Beck}}, \bibinfo {author}
  {\bibfnamefont {J.}~\bibnamefont {Faist}}, \bibinfo {author} {\bibfnamefont
  {S.}~\bibnamefont {Cibella}}, \bibinfo {author} {\bibfnamefont
  {R.}~\bibnamefont {Leoni}}, \bibinfo {author} {\bibfnamefont
  {C.}~\bibnamefont {Reichl}}, \bibinfo {author} {\bibfnamefont
  {C.}~\bibnamefont {Charpentier}}, \ and\ \bibinfo {author} {\bibfnamefont
  {W.}~\bibnamefont {Wegscheider}},\ }\href {\doibase
  10.1103/PhysRevB.90.205309} {\bibfield  {journal} {\bibinfo  {journal} {Phys.
  Rev. B}\ }\textbf {\bibinfo {volume} {90}},\ \bibinfo {pages} {205309}
  (\bibinfo {year} {2014})}\BibitemShut {NoStop}%
\bibitem [{\citenamefont {Scalari}\ \emph {et~al.}(2014)\citenamefont
  {Scalari}, \citenamefont {Maissen}, \citenamefont {Cibella}, \citenamefont
  {Leoni}, \citenamefont {Carelli}, \citenamefont {Valmorra}, \citenamefont
  {Beck},\ and\ \citenamefont {Faist}}]{Scalari2014njp}%
  \BibitemOpen
  \bibfield  {author} {\bibinfo {author} {\bibfnamefont {G.}~\bibnamefont
  {Scalari}}, \bibinfo {author} {\bibfnamefont {C.}~\bibnamefont {Maissen}},
  \bibinfo {author} {\bibfnamefont {S.}~\bibnamefont {Cibella}}, \bibinfo
  {author} {\bibfnamefont {R.}~\bibnamefont {Leoni}}, \bibinfo {author}
  {\bibfnamefont {P.}~\bibnamefont {Carelli}}, \bibinfo {author} {\bibfnamefont
  {F.}~\bibnamefont {Valmorra}}, \bibinfo {author} {\bibfnamefont
  {M.}~\bibnamefont {Beck}}, \ and\ \bibinfo {author} {\bibfnamefont
  {J.}~\bibnamefont {Faist}},\ }\href {\doibase 10.1088/1367-2630/16/3/033005}
  {\bibfield  {journal} {\bibinfo  {journal} {New Journal of Physics}\ }\textbf
  {\bibinfo {volume} {16}},\ \bibinfo {pages} {033005} (\bibinfo {year}
  {2014})}\BibitemShut {NoStop}%
\bibitem [{\citenamefont {Gabbay}\ and\ \citenamefont
  {Brener}(2012)}]{Gabbay2012oe}%
  \BibitemOpen
  \bibfield  {author} {\bibinfo {author} {\bibfnamefont {A.}~\bibnamefont
  {Gabbay}}\ and\ \bibinfo {author} {\bibfnamefont {I.}~\bibnamefont
  {Brener}},\ }\href {\doibase 10.1364/OE.20.006584} {\bibfield  {journal}
  {\bibinfo  {journal} {Optics express}\ }\textbf {\bibinfo {volume} {20}},\
  \bibinfo {pages} {6584} (\bibinfo {year} {2012})}\BibitemShut {NoStop}%
\bibitem [{\citenamefont {Benz}\ \emph
  {et~al.}(2013{\natexlab{a}})\citenamefont {Benz}, \citenamefont {Campione},
  \citenamefont {Liu}, \citenamefont {Montano}, \citenamefont {Klem},
  \citenamefont {Sinclair}, \citenamefont {Capolino},\ and\ \citenamefont
  {Brener}}]{Benz2013oe}%
  \BibitemOpen
  \bibfield  {author} {\bibinfo {author} {\bibfnamefont {A.}~\bibnamefont
  {Benz}}, \bibinfo {author} {\bibfnamefont {S.}~\bibnamefont {Campione}},
  \bibinfo {author} {\bibfnamefont {S.}~\bibnamefont {Liu}}, \bibinfo {author}
  {\bibfnamefont {I.}~\bibnamefont {Montano}}, \bibinfo {author} {\bibfnamefont
  {J.~F.}\ \bibnamefont {Klem}}, \bibinfo {author} {\bibfnamefont {M.~B.}\
  \bibnamefont {Sinclair}}, \bibinfo {author} {\bibfnamefont {F.}~\bibnamefont
  {Capolino}}, \ and\ \bibinfo {author} {\bibfnamefont {I.}~\bibnamefont
  {Brener}},\ }\href {\doibase 10.1364/OE.21.032572} {\bibfield  {journal}
  {\bibinfo  {journal} {Optics Express}\ }\textbf {\bibinfo {volume} {21}},\
  \bibinfo {pages} {32572} (\bibinfo {year} {2013}{\natexlab{a}})}\BibitemShut
  {NoStop}%
\bibitem [{\citenamefont {Benz}\ \emph
  {et~al.}(2013{\natexlab{b}})\citenamefont {Benz}, \citenamefont {Montaño},
  \citenamefont {Klem},\ and\ \citenamefont {Brener}}]{Benz2013apl}%
  \BibitemOpen
  \bibfield  {author} {\bibinfo {author} {\bibfnamefont {A.}~\bibnamefont
  {Benz}}, \bibinfo {author} {\bibfnamefont {I.}~\bibnamefont {Montaño}},
  \bibinfo {author} {\bibfnamefont {J.~F.}\ \bibnamefont {Klem}}, \ and\
  \bibinfo {author} {\bibfnamefont {I.}~\bibnamefont {Brener}},\ }\href
  {\doibase 10.1063/1.4859636} {\bibfield  {journal} {\bibinfo  {journal}
  {Applied Physics Letters}\ }\textbf {\bibinfo {volume} {103}},\ \bibinfo
  {pages} {263116} (\bibinfo {year} {2013}{\natexlab{b}})}\BibitemShut
  {NoStop}%
\bibitem [{\citenamefont {Benz}\ \emph
  {et~al.}(2013{\natexlab{c}})\citenamefont {Benz}, \citenamefont {Campione},
  \citenamefont {Liu}, \citenamefont {Montaño}, \citenamefont {Klem},
  \citenamefont {Allerman}, \citenamefont {Wendt}, \citenamefont {Sinclair},
  \citenamefont {Capolino},\ and\ \citenamefont {Brener}}]{Benz2013nc}%
  \BibitemOpen
  \bibfield  {author} {\bibinfo {author} {\bibfnamefont {A.}~\bibnamefont
  {Benz}}, \bibinfo {author} {\bibfnamefont {S.}~\bibnamefont {Campione}},
  \bibinfo {author} {\bibfnamefont {S.}~\bibnamefont {Liu}}, \bibinfo {author}
  {\bibfnamefont {I.}~\bibnamefont {Montaño}}, \bibinfo {author}
  {\bibfnamefont {J.}~\bibnamefont {Klem}}, \bibinfo {author} {\bibfnamefont
  {A.}~\bibnamefont {Allerman}}, \bibinfo {author} {\bibfnamefont
  {J.}~\bibnamefont {Wendt}}, \bibinfo {author} {\bibfnamefont
  {M.}~\bibnamefont {Sinclair}}, \bibinfo {author} {\bibfnamefont
  {F.}~\bibnamefont {Capolino}}, \ and\ \bibinfo {author} {\bibfnamefont
  {I.}~\bibnamefont {Brener}},\ }\href {\doibase 10.1038/ncomms3882} {\bibfield
   {journal} {\bibinfo  {journal} {Nature Communications}\ }\textbf {\bibinfo
  {volume} {4}} (\bibinfo {year} {2013}{\natexlab{c}}),\
  10.1038/ncomms3882}\BibitemShut {NoStop}%
\bibitem [{\citenamefont {Geiser}\ \emph {et~al.}(2010)\citenamefont {Geiser},
  \citenamefont {Walther}, \citenamefont {Scalari}, \citenamefont {Beck},
  \citenamefont {Fischer}, \citenamefont {Nevou},\ and\ \citenamefont
  {Faist}}]{Geiser2010apl}%
  \BibitemOpen
  \bibfield  {author} {\bibinfo {author} {\bibfnamefont {M.}~\bibnamefont
  {Geiser}}, \bibinfo {author} {\bibfnamefont {C.}~\bibnamefont {Walther}},
  \bibinfo {author} {\bibfnamefont {G.}~\bibnamefont {Scalari}}, \bibinfo
  {author} {\bibfnamefont {M.}~\bibnamefont {Beck}}, \bibinfo {author}
  {\bibfnamefont {M.}~\bibnamefont {Fischer}}, \bibinfo {author} {\bibfnamefont
  {L.}~\bibnamefont {Nevou}}, \ and\ \bibinfo {author} {\bibfnamefont
  {J.}~\bibnamefont {Faist}},\ }\href {\doibase 10.1063/1.3511446} {\bibfield
  {journal} {\bibinfo  {journal} {Applied Physics Letters}\ }\textbf {\bibinfo
  {volume} {97}},\ \bibinfo {pages} {191107} (\bibinfo {year}
  {2010})}\BibitemShut {NoStop}%
\bibitem [{\citenamefont {Dietze}\ \emph {et~al.}(2011)\citenamefont {Dietze},
  \citenamefont {Benz}, \citenamefont {Strasser}, \citenamefont {Unterrainer},\
  and\ \citenamefont {Darmo}}]{Dietze2011oe}%
  \BibitemOpen
  \bibfield  {author} {\bibinfo {author} {\bibfnamefont {D.}~\bibnamefont
  {Dietze}}, \bibinfo {author} {\bibfnamefont {A.}~\bibnamefont {Benz}},
  \bibinfo {author} {\bibfnamefont {G.}~\bibnamefont {Strasser}}, \bibinfo
  {author} {\bibfnamefont {K.}~\bibnamefont {Unterrainer}}, \ and\ \bibinfo
  {author} {\bibfnamefont {J.}~\bibnamefont {Darmo}},\ }\href {\doibase
  10.1364/OE.19.013700} {\bibfield  {journal} {\bibinfo  {journal} {Optics
  express}\ }\textbf {\bibinfo {volume} {19}},\ \bibinfo {pages} {13700}
  (\bibinfo {year} {2011})}\BibitemShut {NoStop}%
\bibitem [{\citenamefont {Geiser}\ \emph {et~al.}(2012)\citenamefont {Geiser},
  \citenamefont {Scalari}, \citenamefont {Castellano}, \citenamefont {Beck},\
  and\ \citenamefont {Faist}}]{Geiser2012apl}%
  \BibitemOpen
  \bibfield  {author} {\bibinfo {author} {\bibfnamefont {M.}~\bibnamefont
  {Geiser}}, \bibinfo {author} {\bibfnamefont {G.}~\bibnamefont {Scalari}},
  \bibinfo {author} {\bibfnamefont {F.}~\bibnamefont {Castellano}}, \bibinfo
  {author} {\bibfnamefont {M.}~\bibnamefont {Beck}}, \ and\ \bibinfo {author}
  {\bibfnamefont {J.}~\bibnamefont {Faist}},\ }\href {\doibase
  10.1063/1.4757611} {\bibfield  {journal} {\bibinfo  {journal} {Applied
  Physics Letters}\ }\textbf {\bibinfo {volume} {101}},\ \bibinfo {pages}
  {141118} (\bibinfo {year} {2012})}\BibitemShut {NoStop}%
\bibitem [{\citenamefont {Dietze}\ \emph {et~al.}(2013)\citenamefont {Dietze},
  \citenamefont {Andrews}, \citenamefont {Klang}, \citenamefont {Strasser},
  \citenamefont {Unterrainer},\ and\ \citenamefont {Darmo}}]{Dietze2013apl}%
  \BibitemOpen
  \bibfield  {author} {\bibinfo {author} {\bibfnamefont {D.}~\bibnamefont
  {Dietze}}, \bibinfo {author} {\bibfnamefont {A.~M.}\ \bibnamefont {Andrews}},
  \bibinfo {author} {\bibfnamefont {P.}~\bibnamefont {Klang}}, \bibinfo
  {author} {\bibfnamefont {G.}~\bibnamefont {Strasser}}, \bibinfo {author}
  {\bibfnamefont {K.}~\bibnamefont {Unterrainer}}, \ and\ \bibinfo {author}
  {\bibfnamefont {J.}~\bibnamefont {Darmo}},\ }\href {\doibase
  10.1063/1.4830092} {\bibfield  {journal} {\bibinfo  {journal} {Applied
  Physics Letters}\ }\textbf {\bibinfo {volume} {103}},\ \bibinfo {pages}
  {201106} (\bibinfo {year} {2013})}\BibitemShut {NoStop}%
\bibitem [{\citenamefont {Shrekenhamer}\ \emph {et~al.}(2011)\citenamefont
  {Shrekenhamer}, \citenamefont {Rout}, \citenamefont {Strikwerda},
  \citenamefont {Bingham}, \citenamefont {Averitt}, \citenamefont {Sonkusale},\
  and\ \citenamefont {Padilla}}]{Shrekenhamer2011oe}%
  \BibitemOpen
  \bibfield  {author} {\bibinfo {author} {\bibfnamefont {D.}~\bibnamefont
  {Shrekenhamer}}, \bibinfo {author} {\bibfnamefont {S.}~\bibnamefont {Rout}},
  \bibinfo {author} {\bibfnamefont {A.~C.}\ \bibnamefont {Strikwerda}},
  \bibinfo {author} {\bibfnamefont {C.}~\bibnamefont {Bingham}}, \bibinfo
  {author} {\bibfnamefont {R.~D.}\ \bibnamefont {Averitt}}, \bibinfo {author}
  {\bibfnamefont {S.}~\bibnamefont {Sonkusale}}, \ and\ \bibinfo {author}
  {\bibfnamefont {W.~J.}\ \bibnamefont {Padilla}},\ }\href {\doibase
  10.1364/OE.19.009968} {\bibfield  {journal} {\bibinfo  {journal} {Opt.
  Express}\ }\textbf {\bibinfo {volume} {19}},\ \bibinfo {pages} {9968}
  (\bibinfo {year} {2011})}\BibitemShut {NoStop}%
\bibitem [{\citenamefont {Pal}\ \emph {et~al.}(2014)\citenamefont {Pal},
  \citenamefont {Valentin}, \citenamefont {Kukharchyk}, \citenamefont {Nong},
  \citenamefont {Parsa}, \citenamefont {Eggeler}, \citenamefont {Ludwig},
  \citenamefont {Jukam},\ and\ \citenamefont {Wieck}}]{Pal2014jpcm}%
  \BibitemOpen
  \bibfield  {author} {\bibinfo {author} {\bibfnamefont {S.}~\bibnamefont
  {Pal}}, \bibinfo {author} {\bibfnamefont {S.~R.}\ \bibnamefont {Valentin}},
  \bibinfo {author} {\bibfnamefont {N.}~\bibnamefont {Kukharchyk}}, \bibinfo
  {author} {\bibfnamefont {H.}~\bibnamefont {Nong}}, \bibinfo {author}
  {\bibfnamefont {A.~B.}\ \bibnamefont {Parsa}}, \bibinfo {author}
  {\bibfnamefont {G.}~\bibnamefont {Eggeler}}, \bibinfo {author} {\bibfnamefont
  {A.}~\bibnamefont {Ludwig}}, \bibinfo {author} {\bibfnamefont
  {N.}~\bibnamefont {Jukam}}, \ and\ \bibinfo {author} {\bibfnamefont {A.~D.}\
  \bibnamefont {Wieck}},\ }\href {\doibase 10.1088/0953-8984/26/50/505801}
  {\bibfield  {journal} {\bibinfo  {journal} {Journal of Physics: Condensed
  Matter}\ }\textbf {\bibinfo {volume} {26}},\ \bibinfo {pages} {505801}
  (\bibinfo {year} {2014})}\BibitemShut {NoStop}%
\bibitem [{\citenamefont {Bastard}\ \emph {et~al.}(1983)\citenamefont
  {Bastard}, \citenamefont {Mendez}, \citenamefont {Chang},\ and\ \citenamefont
  {Esaki}}]{Bastard1983prb}%
  \BibitemOpen
  \bibfield  {author} {\bibinfo {author} {\bibfnamefont {G.}~\bibnamefont
  {Bastard}}, \bibinfo {author} {\bibfnamefont {E.~E.}\ \bibnamefont {Mendez}},
  \bibinfo {author} {\bibfnamefont {L.~L.}\ \bibnamefont {Chang}}, \ and\
  \bibinfo {author} {\bibfnamefont {L.}~\bibnamefont {Esaki}},\ }\href
  {\doibase 10.1103/PhysRevB.28.3241} {\bibfield  {journal} {\bibinfo
  {journal} {Physical Review B}\ }\textbf {\bibinfo {volume} {28}},\ \bibinfo
  {pages} {3241} (\bibinfo {year} {1983})}\BibitemShut {NoStop}%
\bibitem [{\citenamefont {Craig}\ \emph {et~al.}(1996)\citenamefont {Craig},
  \citenamefont {Galdrikian}, \citenamefont {Heyman}, \citenamefont {Markelz},
  \citenamefont {Williams}, \citenamefont {Sherwin}, \citenamefont {Campman},
  \citenamefont {Hopkins},\ and\ \citenamefont {Gossard}}]{Craig1996prl}%
  \BibitemOpen
  \bibfield  {author} {\bibinfo {author} {\bibfnamefont {K.}~\bibnamefont
  {Craig}}, \bibinfo {author} {\bibfnamefont {B.}~\bibnamefont {Galdrikian}},
  \bibinfo {author} {\bibfnamefont {J.~N.}\ \bibnamefont {Heyman}}, \bibinfo
  {author} {\bibfnamefont {A.~G.}\ \bibnamefont {Markelz}}, \bibinfo {author}
  {\bibfnamefont {J.~B.}\ \bibnamefont {Williams}}, \bibinfo {author}
  {\bibfnamefont {M.~S.}\ \bibnamefont {Sherwin}}, \bibinfo {author}
  {\bibfnamefont {K.}~\bibnamefont {Campman}}, \bibinfo {author} {\bibfnamefont
  {P.~F.}\ \bibnamefont {Hopkins}}, \ and\ \bibinfo {author} {\bibfnamefont
  {A.~C.}\ \bibnamefont {Gossard}},\ }\href {\doibase
  10.1103/PhysRevLett.76.2382} {\bibfield  {journal} {\bibinfo  {journal}
  {Physical review letters}\ }\textbf {\bibinfo {volume} {76}},\ \bibinfo
  {pages} {2382} (\bibinfo {year} {1996})}\BibitemShut {NoStop}%
\bibitem [{\citenamefont {Machhadani}\ \emph {et~al.}(2011)\citenamefont
  {Machhadani}, \citenamefont {Tchernycheva}, \citenamefont {Sakr},
  \citenamefont {Rigutti}, \citenamefont {Colombelli}, \citenamefont {Warde},
  \citenamefont {Mietze}, \citenamefont {As},\ and\ \citenamefont
  {Julien}}]{Machhadani2011prb}%
  \BibitemOpen
  \bibfield  {author} {\bibinfo {author} {\bibfnamefont {H.}~\bibnamefont
  {Machhadani}}, \bibinfo {author} {\bibfnamefont {M.}~\bibnamefont
  {Tchernycheva}}, \bibinfo {author} {\bibfnamefont {S.}~\bibnamefont {Sakr}},
  \bibinfo {author} {\bibfnamefont {L.}~\bibnamefont {Rigutti}}, \bibinfo
  {author} {\bibfnamefont {R.}~\bibnamefont {Colombelli}}, \bibinfo {author}
  {\bibfnamefont {E.}~\bibnamefont {Warde}}, \bibinfo {author} {\bibfnamefont
  {C.}~\bibnamefont {Mietze}}, \bibinfo {author} {\bibfnamefont {D.~J.}\
  \bibnamefont {As}}, \ and\ \bibinfo {author} {\bibfnamefont {F.~H.}\
  \bibnamefont {Julien}},\ }\href {\doibase 10.1103/PhysRevB.83.075313}
  {\bibfield  {journal} {\bibinfo  {journal} {Physical Review B}\ }\textbf
  {\bibinfo {volume} {83}} (\bibinfo {year} {2011}),\
  10.1103/PhysRevB.83.075313}\BibitemShut {NoStop}%
\bibitem [{cst()}]{cst}%
  \BibitemOpen
  \href@noop {} {\enquote {\bibinfo {title}
  {https://www.cst.com/applications},}\ }\BibitemShut {NoStop}%
\bibitem [{\citenamefont {Shelton}\ \emph {et~al.}(2010)\citenamefont
  {Shelton}, \citenamefont {Peters}, \citenamefont {Sinclair}, \citenamefont
  {Brener}, \citenamefont {Warne}, \citenamefont {Basilio}, \citenamefont
  {Coffey},\ and\ \citenamefont {Boreman}}]{Shelton2010oe}%
  \BibitemOpen
  \bibfield  {author} {\bibinfo {author} {\bibfnamefont {D.~J.}\ \bibnamefont
  {Shelton}}, \bibinfo {author} {\bibfnamefont {D.~W.}\ \bibnamefont {Peters}},
  \bibinfo {author} {\bibfnamefont {M.~B.}\ \bibnamefont {Sinclair}}, \bibinfo
  {author} {\bibfnamefont {I.}~\bibnamefont {Brener}}, \bibinfo {author}
  {\bibfnamefont {L.~K.}\ \bibnamefont {Warne}}, \bibinfo {author}
  {\bibfnamefont {L.~I.}\ \bibnamefont {Basilio}}, \bibinfo {author}
  {\bibfnamefont {K.~R.}\ \bibnamefont {Coffey}}, \ and\ \bibinfo {author}
  {\bibfnamefont {G.~D.}\ \bibnamefont {Boreman}},\ }\href {\doibase
  10.1364/OE.18.001085} {\bibfield  {journal} {\bibinfo  {journal} {Opt.
  Express}\ }\textbf {\bibinfo {volume} {18}},\ \bibinfo {pages} {1085}
  (\bibinfo {year} {2010})}\BibitemShut {NoStop}%
\bibitem [{\citenamefont {Benz}\ \emph {et~al.}(2015)\citenamefont {Benz},
  \citenamefont {Campione}, \citenamefont {Klem}, \citenamefont {Sinclair},\
  and\ \citenamefont {Brener}}]{Benz2015nl}%
  \BibitemOpen
  \bibfield  {author} {\bibinfo {author} {\bibfnamefont {A.}~\bibnamefont
  {Benz}}, \bibinfo {author} {\bibfnamefont {S.}~\bibnamefont {Campione}},
  \bibinfo {author} {\bibfnamefont {J.~F.}\ \bibnamefont {Klem}}, \bibinfo
  {author} {\bibfnamefont {M.~B.}\ \bibnamefont {Sinclair}}, \ and\ \bibinfo
  {author} {\bibfnamefont {I.}~\bibnamefont {Brener}},\ }\href {\doibase
  10.1021/nl504815c} {\bibfield  {journal} {\bibinfo  {journal} {Nano Letters}\
  }\textbf {\bibinfo {volume} {15}},\ \bibinfo {pages} {1959} (\bibinfo {year}
  {2015})}\BibitemShut {NoStop}%
\bibitem [{\citenamefont {Snider}\ \emph {et~al.}(1990)\citenamefont {Snider},
  \citenamefont {Tan},\ and\ \citenamefont {Hu}}]{Snider1990jap}%
  \BibitemOpen
  \bibfield  {author} {\bibinfo {author} {\bibfnamefont {G.~L.}\ \bibnamefont
  {Snider}}, \bibinfo {author} {\bibfnamefont {I.-H.}\ \bibnamefont {Tan}}, \
  and\ \bibinfo {author} {\bibfnamefont {E.~L.}\ \bibnamefont {Hu}},\ }\href
  {\doibase 10.1063/1.346443} {\bibfield  {journal} {\bibinfo  {journal}
  {Journal of Applied Physics}\ }\textbf {\bibinfo {volume} {68}},\ \bibinfo
  {pages} {2849} (\bibinfo {year} {1990})}\BibitemShut {NoStop}%
\bibitem [{\citenamefont {Ando}(1976)}]{Ando1976prb}%
  \BibitemOpen
  \bibfield  {author} {\bibinfo {author} {\bibfnamefont {T.}~\bibnamefont
  {Ando}},\ }\href {\doibase 10.1103/PhysRevB.13.3468} {\bibfield  {journal}
  {\bibinfo  {journal} {Phys. Rev. B}\ }\textbf {\bibinfo {volume} {13}},\
  \bibinfo {pages} {3468} (\bibinfo {year} {1976})}\BibitemShut {NoStop}%
\bibitem [{\citenamefont {Wieck}\ \emph {et~al.}(1989)\citenamefont {Wieck},
  \citenamefont {Thiele}, \citenamefont {Merkt}, \citenamefont {Ploog},
  \citenamefont {Weimann},\ and\ \citenamefont {Schlapp}}]{Wieck1989prb}%
  \BibitemOpen
  \bibfield  {author} {\bibinfo {author} {\bibfnamefont {A.~D.}\ \bibnamefont
  {Wieck}}, \bibinfo {author} {\bibfnamefont {F.}~\bibnamefont {Thiele}},
  \bibinfo {author} {\bibfnamefont {U.}~\bibnamefont {Merkt}}, \bibinfo
  {author} {\bibfnamefont {K.}~\bibnamefont {Ploog}}, \bibinfo {author}
  {\bibfnamefont {G.}~\bibnamefont {Weimann}}, \ and\ \bibinfo {author}
  {\bibfnamefont {W.}~\bibnamefont {Schlapp}},\ }\href {\doibase
  10.1103/PhysRevB.39.3785} {\bibfield  {journal} {\bibinfo  {journal} {Phys.
  Rev. B}\ }\textbf {\bibinfo {volume} {39}},\ \bibinfo {pages} {3785}
  (\bibinfo {year} {1989})}\BibitemShut {NoStop}%
\bibitem [{\citenamefont {Wieck}\ \emph {et~al.}(1988)\citenamefont {Wieck},
  \citenamefont {Bollweg}, \citenamefont {Merkt}, \citenamefont {Weimann},\
  and\ \citenamefont {Schlapp}}]{Wieck1988prb}%
  \BibitemOpen
  \bibfield  {author} {\bibinfo {author} {\bibfnamefont {A.~D.}\ \bibnamefont
  {Wieck}}, \bibinfo {author} {\bibfnamefont {K.}~\bibnamefont {Bollweg}},
  \bibinfo {author} {\bibfnamefont {U.}~\bibnamefont {Merkt}}, \bibinfo
  {author} {\bibfnamefont {G.}~\bibnamefont {Weimann}}, \ and\ \bibinfo
  {author} {\bibfnamefont {W.}~\bibnamefont {Schlapp}},\ }\href {\doibase
  10.1103/PhysRevB.38.10158} {\bibfield  {journal} {\bibinfo  {journal} {Phys.
  Rev. B}\ }\textbf {\bibinfo {volume} {38}},\ \bibinfo {pages} {10158}
  (\bibinfo {year} {1988})}\BibitemShut {NoStop}%
\bibitem [{\citenamefont {Padilla}\ \emph {et~al.}(2006)\citenamefont
  {Padilla}, \citenamefont {Taylor}, \citenamefont {Highstrete}, \citenamefont
  {Lee},\ and\ \citenamefont {Averitt}}]{Padilla2006prl}%
  \BibitemOpen
  \bibfield  {author} {\bibinfo {author} {\bibfnamefont {W.~J.}\ \bibnamefont
  {Padilla}}, \bibinfo {author} {\bibfnamefont {A.~J.}\ \bibnamefont {Taylor}},
  \bibinfo {author} {\bibfnamefont {C.}~\bibnamefont {Highstrete}}, \bibinfo
  {author} {\bibfnamefont {M.}~\bibnamefont {Lee}}, \ and\ \bibinfo {author}
  {\bibfnamefont {R.~D.}\ \bibnamefont {Averitt}},\ }\href {\doibase
  10.1103/PhysRevLett.96.107401} {\bibfield  {journal} {\bibinfo  {journal}
  {Physical Review Letters}\ }\textbf {\bibinfo {volume} {96}} (\bibinfo {year}
  {2006}),\ 10.1103/PhysRevLett.96.107401}\BibitemShut {NoStop}%
\bibitem [{\citenamefont {Ginn}\ \emph {et~al.}(2009)\citenamefont {Ginn},
  \citenamefont {Shelton}, \citenamefont {Krenz}, \citenamefont {Lail},\ and\
  \citenamefont {Boreman}}]{Ginn2009jap}%
  \BibitemOpen
  \bibfield  {author} {\bibinfo {author} {\bibfnamefont {J.}~\bibnamefont
  {Ginn}}, \bibinfo {author} {\bibfnamefont {D.}~\bibnamefont {Shelton}},
  \bibinfo {author} {\bibfnamefont {P.}~\bibnamefont {Krenz}}, \bibinfo
  {author} {\bibfnamefont {B.}~\bibnamefont {Lail}}, \ and\ \bibinfo {author}
  {\bibfnamefont {G.}~\bibnamefont {Boreman}},\ }\href {\doibase
  10.1063/1.3093698} {\bibfield  {journal} {\bibinfo  {journal} {Journal of
  Applied Physics}\ }\textbf {\bibinfo {volume} {105}},\ \bibinfo {pages}
  {074304} (\bibinfo {year} {2009})}\BibitemShut {NoStop}%
\bibitem [{\citenamefont {Mooney}(1990)}]{Mooney1990jap}%
  \BibitemOpen
  \bibfield  {author} {\bibinfo {author} {\bibfnamefont {P.~M.}\ \bibnamefont
  {Mooney}},\ }\href {\doibase 10.1063/1.345628} {\bibfield  {journal}
  {\bibinfo  {journal} {Journal of Applied Physics}\ }\textbf {\bibinfo
  {volume} {67}},\ \bibinfo {pages} {R1} (\bibinfo {year} {1990})}\BibitemShut
  {NoStop}%
\bibitem [{\citenamefont {Chen}\ \emph {et~al.}(2006)\citenamefont {Chen},
  \citenamefont {Padilla}, \citenamefont {Zide}, \citenamefont {Gossard},
  \citenamefont {Taylor},\ and\ \citenamefont {Averitt}}]{Chen2006nl}%
  \BibitemOpen
  \bibfield  {author} {\bibinfo {author} {\bibfnamefont {H.-T.}\ \bibnamefont
  {Chen}}, \bibinfo {author} {\bibfnamefont {W.~J.}\ \bibnamefont {Padilla}},
  \bibinfo {author} {\bibfnamefont {J.~M.}\ \bibnamefont {Zide}}, \bibinfo
  {author} {\bibfnamefont {A.~C.}\ \bibnamefont {Gossard}}, \bibinfo {author}
  {\bibfnamefont {A.~J.}\ \bibnamefont {Taylor}}, \ and\ \bibinfo {author}
  {\bibfnamefont {R.~D.}\ \bibnamefont {Averitt}},\ }\href {\doibase
  10.1038/nature05343} {\bibfield  {journal} {\bibinfo  {journal} {Nature}\
  }\textbf {\bibinfo {volume} {444}},\ \bibinfo {pages} {597} (\bibinfo {year}
  {2006})}\BibitemShut {NoStop}%
\bibitem [{\citenamefont {Ando}(1977)}]{Ando1977ssc}%
  \BibitemOpen
  \bibfield  {author} {\bibinfo {author} {\bibfnamefont {T.}~\bibnamefont
  {Ando}},\ }\href {\doibase http://dx.doi.org/10.1016/0038-1098(77)91495-8}
  {\bibfield  {journal} {\bibinfo  {journal} {Solid State Communications}\
  }\textbf {\bibinfo {volume} {21}},\ \bibinfo {pages} {133 } (\bibinfo {year}
  {1977})}\BibitemShut {NoStop}%
\end{thebibliography}%
\noindent{{\bf Acknowledgements:}}
The work is financially supported by the BMBF Quantum communication program - Q.com-H 16KIS0109 and BMBF-QUIMP 16BQ1062 as well as Mercur Pr-2013-0001. The authors would also like to acknowledge the DFH/UFA CDFA-05-06 Nice-Bochum and the RUB Research School. Additionally, S.P. and A.D.W. would like to acknowledge IMPRS-SurMat, MPIE D\"{u}sseldorf.\\

\noindent{{\bf Author Contributions:}}
S.P. and S.M. designed and fabricated the metamaterials and performed the FDTD simulations. S.P. performed electrical measurements and wrote the manuscript; S.P. and N.K. performed the FTIR measurements; H.N. and S.M. developed the set-up for time domain spectroscopy and performed the measurements; N.K. performed the 2DEG simulations. S.R.V. and A.L. designed and grew the samples; S.S. took the SEM pictures; C.B. and U.K. prepared the chrome-masks for metamaterials; A.D.W. and N.J. supervised the experiments and analysed the results. All authors have discussed the results and commented on the manuscript at all stages.
\end{document}